\documentclass[10pt,reqno]{article}
\usepackage{graphicx}
\usepackage{authblk}
\usepackage{natbib}
\usepackage{tablefootnote}
\usepackage[export]{adjustbox}

\usepackage{amssymb,amsthm,amsmath}
\usepackage{xcolor,paralist,hyperref,titlesec,fancyhdr,etoolbox}

\hypersetup{ colorlinks=true, linkcolor=black, filecolor=black, urlcolor=black }

\begin{document}

\title{Crystal Eye: all sky MeV monitor with high precision real-time localization}

\author[a,b]{R.~Aloisio}
\author[a,b]{U.~Atalay}
\author[a,b]{B.~Banerjee}
\author[a,b]{F.~C.~T.~Barbato}
\author[c,d]{E.~Bissaldi}
\author[a,b]{M.~Branchesi}
\author[e]{F.~Capitanio}
\author[a,b]{E.~Casilli}
\author[f,g]{R.~Colalillo}
\author[a,b]{I.~De Mitri}
\author[a,b]{A.~De Santis}
\author[a,b]{A.~Di Giovanni}
\author[h]{M.~Fernandez Alonso}
\author[a,b]{G.~Fontanella}
\author[d]{F.~Gargano}
\author[f,g]{F.~Garufi}
\author[f,g]{F.~Guarino}
\author[a,b]{D.~Kyratzis}
\author[a,b]{H.~Lima}
\author[i,d]{F.~Loparco}
\author[j]{F.~Longo}
\author[j]{R.~Martinelli}
\author[k]{T.~Montaruli}
\author[a,b]{G.~Oganesyan}
\author[l]{J.~Rico}
\author[a,b]{F.~Santoliquido}
\author[a,b]{R.~Sarkar\thanks{Corresponding author: ritabrata.sarkar.gssi.it}} 
\author[a,b]{P.~Savina}
\author[a,b]{I.~Siddique}
\author[a,b]{A.~Smirnov}
\author[f,g]{M.~Tambone}
\author[e]{A.~Tarana}
\author[k]{A.~Tykhonov}
\author[f,g]{L.~Valore}
\author[j]{A.~A.~Vigliano}
\author[m]{L.~Wu}

\date{}

\affil[a]{Gran Sasso Science Institute, L’Aquila, Italy}
\affil[b]{Istituto Nazionale di Fisica Nucleare --- Laboratori Nazionali del Gran Sasso, L’Aquila, Italy}
\affil[c]{Politecnico di Bari, Bari, Italy}
\affil[d]{Istituto Nazionale di Fisica Nucleare - Sezione di Bari, Bari, Italy}
\affil[e]{INAF Istituto di Astrofisica e Planetologia Spaziali, Roma, Italy}
\affil[f]{Università degli Studi di Napoli Federico II - Dipartimento di Fisica `Ettore Pancini', Napoli, Italy}
\affil[g]{Istituto Nazionale di Fisica Nucleare - Sezione di Napoli, Napoli, Italy}
\affil[h]{Université Libre de Bruxelles, Science Faculty CP230, Brussels, Belgium}
\affil[i]{Università degli Studi di Bari Aldo Moro, Bari, Italy}
\affil[j]{Università degli Studi di Trieste, Trieste, Italy}
\affil[k]{DPNC, Université de Genève, Geneva, Switzerland}
\affil[l]{Institut de Física d’Altes Energies, The Barcelona Institute of Science and Technology, Barcelona, Spain}
\affil[m]{Institute of Deep Space Sciences, Deep Space Exploration Laboratory, Hefei, China}

\maketitle


\begin{abstract}
Crystal~Eye is a space-based all-sky monitor optimized for the autonomous detection
and localization of transients in the 10 keV to 30 MeV energy range, a region where
extensive observations and monitoring of various astrophysical phenomena are required.
By focusing on the operating environment and its impact on the observation process,
we optimized the detector design and assessed its scientific potential. We explored
the use of novel techniques to achieve the science goals of the experiment. We assumed
the orbit of a potential future mission at approximately 550 km altitude near the
equatorial region with a 20$^{\circ}$ inclination. In such an orbit, the main background
contributions for this kind of detector are from different particles and radiation of
cosmic origin and secondaries produced by their interaction in the Earth's atmospheric
and geomagnetic environment. We studied the response of Crystal~Eye detector in this
background environment, using the Geant4 Monte Carlo simulation toolkit. We also
calculated other detector performance parameters to estimate its scientific capabilities.
The effective area and efficiency of the detector are calculated for low energy
$\gamma$-ray sources and used to estimate its sensitivity to short-duration transient
sources. The calculation shows a better effective area and sensitivity
by several factors compared to existing instruments of similar type. A method is also
developed and discussed to estimate the online transient-localization performance of
the detector, suggesting a better localization precision by about an order of magnitude
than those typically reported by existing $\gamma$-ray monitors. We present here the
simulation study and results of an innovative detector design concept that can make
a significant contribution in the multi-messenger era. Moreover, this study can be
useful as a technical reference for similar future experiments.
\end{abstract} 


\section{Introduction}
\label{sec:intro}

Monitoring the sky in the MeV energy domain is essential to uncover the nature of
explosive phenomena in the Universe. In the scale of milliseconds to hours, bursts
of MeV radiation are observed from various physical sources: from the thunderstorms
at Earth to the most violent explosions triggered by the death of massive stars and
mergers of compact objects. Given that the Universe is almost transparent to the MeV
$\gamma$-rays, we have access to the most distant transients, such as gamma-ray bursts
(GRBs). For the last three decades, the prompt localization of GRBs and their
multi-wavelength follow-up have allowed us to identify two distinct progenitors, namely
collapsars and coalescence of neutron stars. In 2017, a joint detection of the
gravitational wave (GW) event from the merger of binary neutron stars (GW170817) and
GRB170817A initiated a new multi-messenger era with GWs and electromagnetic (EM) radiation
\citep{abbott_2017, goldstein_2017}. In the upcoming decade, GW astronomy is expected
to gain more precision, allowing for a sensitivity increase of more than one order of
magnitude from few Hz to kHz frequency range (Einstein Telescope \citep{punturo_2010}
and Cosmic Explorer \citep{reitze_2019}). In return, we expect a dramatic increase in
the rate of joint GW and EM observations, from a few events per year (advanced design of
LIGO-Virgo KAGRA \citep{ligo_2015, virgo_2015}, from late 2027) to several hundreds
per year \citep{ronchini_2022}.

The rapidly growing field of multi-messenger astronomy with GWs requires advancement in
the EM facilities. On the one hand, currently operating MeV telescopes are already over
their expected lifetime and could be decommissioned in a few years. On the other hand,
the new MeV instrumentation should address the current limitations of MeV monitors. The
Neil Gehrels Swift Observatory \citep{gehrels_2004} has a unique capacity for fast arcmin
localization of GRBs while operating in the hard X-ray regime (15--350 keV). In contrast,
the gamma-ray burst monitor (GBM) onboard the Fermi gamma-ray space telescope \citep{meegan_2009}
faces problems of huge localization uncertainties ($\sim$ 10--100 square degrees) while
characterizing GRBs in the broad range of 8 keV to 40 MeV and even beyond (> 100 MeV,
Large Area Telescope).

Crystal~Eye (CE) is designed as an all-sky monitor sensitive to photons of energy of 10 keV
to 30 MeV \citep{barbato_2019}. The overall shape and active media arrangement of CE are
specifically designed to maximize performance and optimize scientific observations. CE stands
out for its broader energy coverage, higher sensitivity, and localization accuracy over a full-sky
field of view (FoV). The autonomous all-sky monitoring and localization capability on board
makes it uniquely suited for real-time transient and multi-messenger astrophysics in the MeV
regime. The higher sensitivity of CE allows us to discover and/or characterize several classes
of transients, including stellar flares, novae, magnetar flares, relativistic shock breakout
signals, extragalactic jetted objects, and more.

The detector will make use of some of the latest photon detection technologies,
including silicon photomultiplier (SiPM) and novel scintillating materials, to achieve
its scientific goals. Among the primary scientific targets of the instrument there are
GRBs, the counterparts of gravitational waves and other transient neutrino
sources, accreting systems, supernovae, and particular $\gamma$-ray emission lines from
the nuclear reactions in exotic astrophysical sources.

Observation of these phenomena requires excellent instrumental performance in various
aspects, while dealing with a complex radiation and particle background environment
that dominates the energy range. In addition to the predominating cosmic diffused and
albedo photon background, neutrons from the atmospheric interaction of high-energy
cosmic rays, primary and secondary protons (trapped or free), and e$^-$/e$^+$ trapped
in Earth's magnetic field are among the particles potentially affecting the
detection process and detector sensitivity. The omnidirectional occurrence of
the astrophysical transient phenomena require the detector to have a good localization
capability with high angular resolution while maintaining the wide FoV. All these
aspects and requirements have been considered to conceptualize the detector design.
The best performance can be obtained with a constellation of 3 hemispherical detector
modules at optimal orbital placement. The combined observation of these modules will
allow for the all-sky coverage and a zenith-uniform response efficiency.

In this study, the Geant4 simulation toolkit\footnote{version 11.2.1} \citep{agostinelli_2003}
is used to develop a detailed geometry of a single CE module and simulate the interaction
of different particles and photons in the detector. Thus, the response of the instrument
to its particular background environment can be estimated along with other performance
parameters including the effective area, its sensitivity to persistent and transient
sources, and the localization power. Apart from the external background due to the orbital
radiation environment, there may be some (significant) internal source of background, such
as the presence of radioactive isotopes in the scintillator-crystal material itself,
depending on the type of material in use. In this study, we also considered these effects
for the optimization of the instrument design and the estimation of the instrument performance.

This paper is structured as follows. In Section~\ref{sec:detDesc}, the overall structure
of the detector and its components are described. The general properties of the detector
and discussions on the simulation procedure are presented in Section~\ref{sec:detProp}.
In Section~\ref{sec:bkg}, we give details of the background estimation for the detector
in its operational environment. The sensitivity of the detector and its response to the
target sources are discussed in Section~\ref{sec:sensitivity}, and the source localization
of the detector is presented in Section~\ref{sec:srcloc}. Finally, Section~\ref{sec:conc}
provides a summary of the conclusions and outlines future directions.

\section{The Crystal~Eye detector}
\label{sec:detDesc}

The design of a space-bound detector is unavoidably a compromise between size, weight,
and performance, where materials and hardware play a key role. Gamma-ray detection
technologies have experienced major advances over the past decades, with the introduction of
efficient scintillating materials and affordable and compact single-photon sensitive
devices like SiPM. CE is mainly intended to operate as an all-sky monitor, with a design
that builds up its concept on these technological advancements and therefore integrates
optimized detection efficiency, sensitivity, and FoV, with autonomous real-time onboard
localization of high-energy transients, supporting low-latency alerts for multi-wavelength
and multi-messenger follow-up.

\subsection{Instrument description}
\label{ssec:instru}

\begin{figure}[htp]
  \centering
  \includegraphics[width=0.5\hsize]{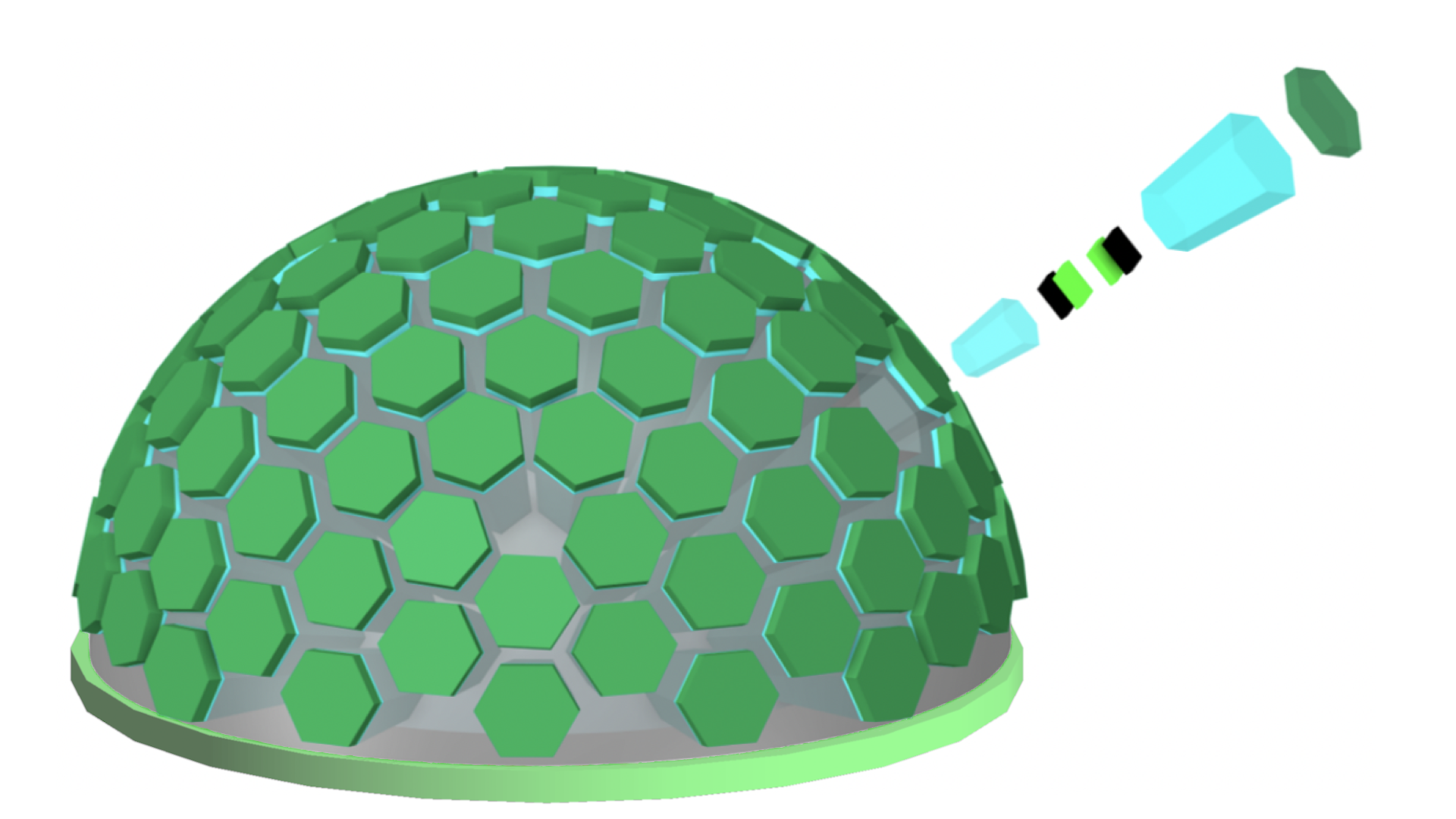}
  \caption{Crystal Eye detector exploded schematic view. Each pixel consists of two
  crystals arranged in two concentric layers with the signal readout components in
  between them. An outer layer of segmented plastic scintillators on the hemisphere
  and a disc of same material at the bottom serve as the veto layers.}
  \label{fig:dome}
\end{figure}

Each CE module consists of a dome-shaped structure with an overall diameter of $\sim$ 32 cm,
composed of 112 scintillator crystal units or pixels (made of two crystals) that cover a
2$\pi$ FoV locally (see Fig.~\ref{fig:dome}). The current design employs SiPMs, compact
photosensitive devices with high detection efficiency, low power consumption, and
insensitivity to magnetic fields. The pixel layout is arranged in two layers, designed to
maximize the surface coverage with optimal granularity and high $\gamma$-ray detection
efficiency between 10 keV and 3 MeV for the outer layer, which extend up to 30 MeV considering
both layers (see Fig.~\ref{fig:crysEff}).

A segmented layer of plastic scintillator at the top of each pixel covers the entire upper
surface of the detector dome to perform two tasks: veto and hard X-ray detector. Working
in anti-coincidence with the crystals, it vetoes and tags the charged particles, but can
also be used to detect the hard X-rays. Another layer of disk-shaped plastic scintillator
is placed at the bottom of the dome to discriminate the particles coming from the bottom
and to identify the non-contained shower events in the detector. Different trigger logics
can be set to suppress or minimize the background by considering the amount and topology
of the energy depositions in the detector crystals and veto layers by different particles.

\subsection{The pixel}
\label{ssec:pixel}
Each pixel of the detector consists of two scintillating crystals (hereafter referred
as calorimeter or CAL crystals) forming the two concentric layers of the detector
(see Fig.~\ref{fig:dome}). The CAL crystals are covered by the veto layers operating
in the anti-coincidence mode (hereafter referred as ACD), for discrimination of charged
particles from photons and partial energy depositions. The granularity of the pixels
in the detector provides the ability to speculate on the direction of the detected
events. It also allows us to use simpler electronics, while simultaneously providing
thermal and radiation insulation for the electronic components. Currently, different
studies are being performed to determine the type of crystal to be used, to achieve a
realistic and sustainable mechanical design, and to optimize the signal readout system.
Among the crystal candidates we consider, LYSO: a lutetium-based scintillation crystal,
and GAGG: gadolinium aluminum gallium garnet, which possesses to some extent similar
properties as highlighted in Table~\ref{tab:crys}.

\begin{table}[htp]
  \centering
  \begin{tabular}{l c c}
    \hline
    Properties                                  & LYSO          & GAGG \\
    \hline
        Density [g\,cm$^{-3}$]                  & 7.25          & 6.60 \\
        Refractive index                        & 1.82          & 1.91 \\
        Light output [photons\,MeV$^{-1}$]      & 30\,000       & 30\,000 \\
        Wavelength of emission peak [nm]        & 420           & 520 \\
        Decay constant [ns]                     & 40            & 50 \\
        Energy resolution [\% @662 keV]         & 10.9          & 7.0 \\
    \hline
  \end{tabular}
  \caption{General properties of the scintillator crystal materials LYSO and GAGG.\tablefootnote{https://www.epic-crystal.com}}
  \label{tab:crys}
\end{table}

Both LYSO \citep{cooke_2000} and GAGG \citep{kamada_2012} are cerium-doped scintillator
crystals that have been developed in recent times and have several advantages over
commonly used scintillator materials. They exhibit a high light yield and a fast
decay time, which enables the instrument to discriminate the temporal features in the
millisecond timescale. They also have high densities, which naturally leads to a compact
detector design. One peculiar property of LYSO crystals, their intrinsic radiation,
due to lutetium, can in principle be used to self-calibrate the detector in energy during
its operation in orbit. However, this same feature could potentially become a source of
significant background in the CE's operative energy range and therefore must be well
characterized and/or suppressed with appropriate selection cuts and design consideration.
While GAGG is essentially free from the internal background radiation, it has slightly
lower density, shows some non-linearity in the light-yield for different energy
depositions, and affects the budget of the experiment.

\begin{figure}[htp]
  \centering
  \includegraphics[width=0.59\hsize,valign=c]{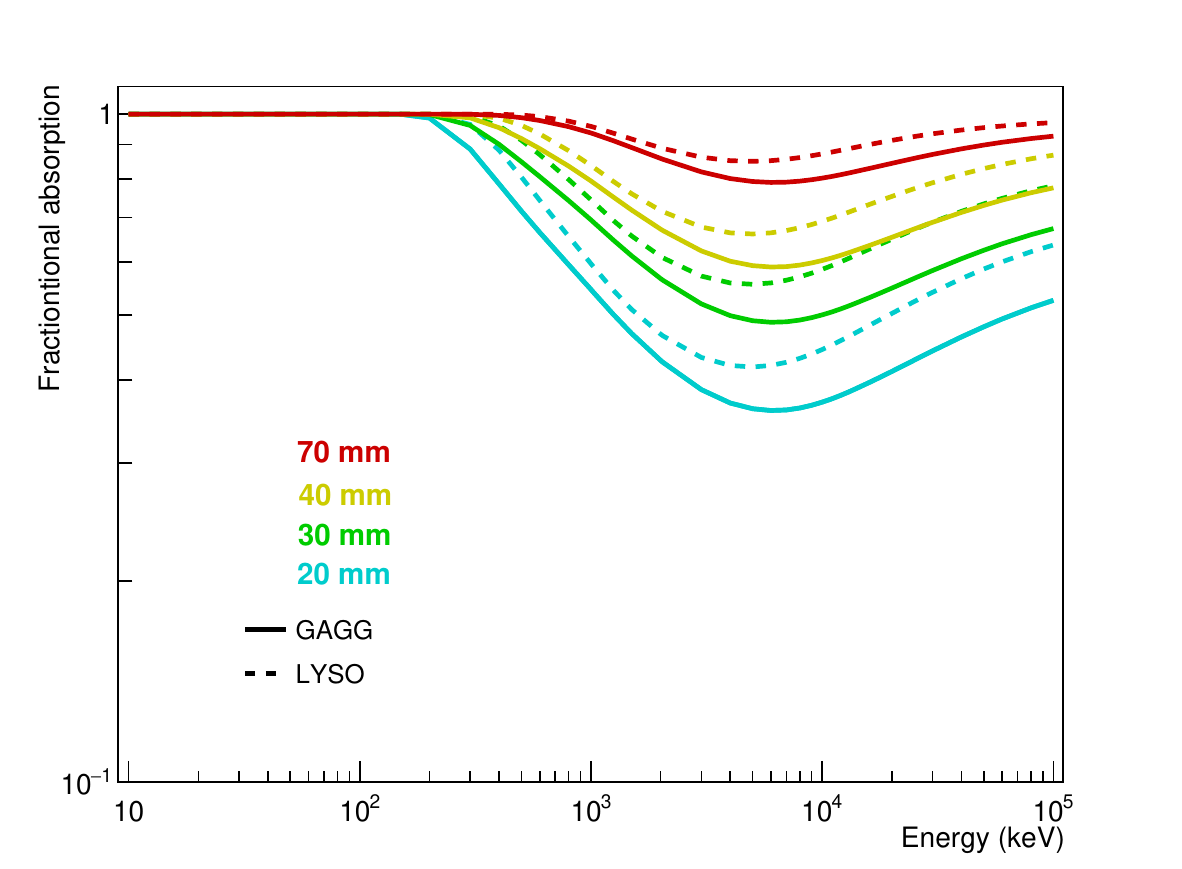}
  \includegraphics[width=0.15\hsize,valign=c]{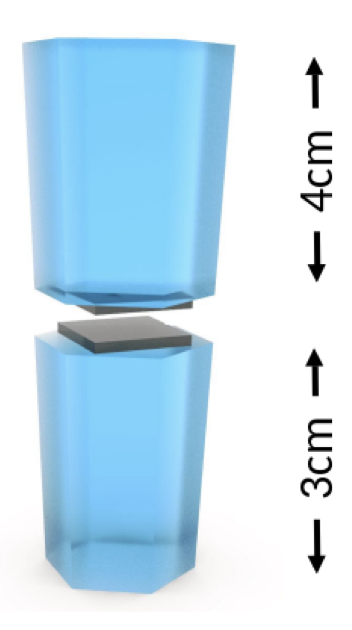}
  \caption{LYSO and GAGG absorption efficiency for different material depths.
  To the right, schematic position and dimension of the crystals used in the pixels
  are shown.}
  \label{fig:crysEff}
\end{figure}

Figure~\ref{fig:crysEff} shows the absorption efficiency of LYSO and GAGG as a function
of energy for different material depths. Considering the absorption efficiency for
different material thicknesses, the granularity requirement for the direction localization
of the photon sources, the weight of the instrument, and other important parameters,
the top and bottom crystals can be optimally dimensioned as follows. The top crystals
have a trunk pyramidal form with a hexagonal base and height of 40 mm. The bottom crystals
follow the same pyramid with a height of 30 mm. The depth of the top pixels results in an
absorption efficiency above 65\% (LYSO) up to 10 MeV, while it reaches above 85\% when
considering both layers. This quantity directly affects the overall efficiency of the
detector and (partially) determines the energy range where the instrument will be sensitive.
The other deciding factors for the energy range are electronics, background, etc. The
thickness of the segmented hexagonal top ACDs is 5 mm, while that of the ACD at the bottom
is 10 mm.

Taking into consideration the absorption efficiency, scintillation properties, and
financial aspects, the primary choice for the crystal material is LYSO. However,
as we show in more detail in Section~\ref{sec:bkg}, a major drawback for LYSO is the
presence of a significant radiation background due to the internal radiation of the
radioactive isotope present in the crystal material. We tried to optimize the design
structure considering a combination of LYSO and GAGG. In this work, we carry out
a comparative study of different possible detector configurations: one using LYSO
for all CAL crystals (``LYSO configuration''), and the other, where the top CAL
crystals are GAGG and the bottom are LYSO (``LYSO\,+\,GAGG configuration''). The
possibility of a third configuration using only GAGG crystals (``GAGG configuration''),
which is essentially free from the intrinsic radioactive background, is also discussed
to compare some important performance parameters of the detector.

\section{Estimation of detector properties}
\label{sec:detProp}

We calculated the general performance properties of the detector such as the
effective area and efficiency, using a Monte Carlo simulation, and then derived
the effective detector background from the operative radiation environment of the
experiment. The estimated detector background is subsequently used to derive the
sensitivity of the detector for different types of astrophysical sources. The
localization capability of the instrument is another crucial aspect of the experiment
which is discussed in Section~\ref{sec:srcloc}. In this section, we describe
the simulation procedure and selection cuts applied on the events, which are used
to compute the detector properties, background, and other parameters such as
sensitivity and source localization precision.

\subsection{Simulation procedure}
\label{ssec:simu}

We considered a detailed geometrical model of the detector for the simulation using
the Geant4 simulation toolkit. We implemented the geometrical description of the
detector, as shown in Fig.~\ref{fig:dome} and discussed in Section~\ref{sec:detDesc},
using computer-aided design (CAD), including the CAL crystals, the ACD layers. We also
considered a simplified internal structure made of aluminum that holds the crystals.
However, the SiPMs for signal readout, other electronic components and cables, and
the satellite structure to host the detector are not considered for the time being.

To simulate the interaction of different particles and radiation in detector materials,
we considered a customized list of physical processes available in the Geant4 toolkit.
G4EmStandard-Physics module is applied for the EM processes, whereas the FTFP\_BERT
physics list is used to address the hadronic interactions. G4Radioactivation and
G4DecayPhysics are also activated to take care of radioactivity and decay of unstable
particles, respectively. Some additional elastic and inelastic processes are also
considered for hadronic and nuclear interactions. The physics list is also carefully
curated to take into account the decay of radioactive isotopes in LYSO crystals.

\subsection{Trigger conditions and selection cuts}
\label{ssec:trig}

In order to select high-quality events, maximize background rejection, and avoid
electronic noise, some threshold cuts are required for trigger and event selection.
The threshold cuts are intended to emulate a real experimental environment enabling
electronic noise suppression. The applied threshold cut values to suppress the
electronic noise are 7 keV and 30 keV for each ACD and CAL crystals, respectively.
These values are obtained from preliminary laboratory test results of detector-readout
optimization. Studies are on going to reduce these threshold values in order to achieve
the lower limit of the proposed energy range of the experiment. Events with energy
deposition below these thresholds are discarded. In addition, some basic selection
cuts are also applied, in order to select the events of interest for the analysis
(both for all the detector configurations). The following selection conditions are
applied as the ``basic trigger'' in this study:
\begin{itemize}
  \item total deposited energy in the top ACD layer < 200 keV: selecting photons over
  particle events (electrons, protons, etc.);
  \item no energy deposition signal from the bottom ACD: to remove albedo backgrounds
  and excluding the non-contained events depositing energy in the calorimeter;
  \item deposited energy in the upper layer CAL crystals > lower layer CAL crystals:
  to ensure proper energy deposition in the calorimeter for those events coming from
  the upper side of the detector.
\end{itemize}

In addition, another trigger condition was considered, in particular for the
LYSO\,+\,GAGG configuration, depending on the topology of the energy distribution in
the CAL crystals and is called the ``topological trigger''. In this condition:
\begin{itemize}
  \item the maximum amount of energy deposition in a single CAL crystal for an event
  must belong to the top layer;
  \item at least one of the other CAL crystals with energy deposition (if any) should
  be at the immediate vicinity of that with maximum energy deposition (may belong to
  either layer);
  \item the combined energy deposition in the cluster of CAL crystals surrounding the
  one with the maximum energy deposition is more than 50\% of the total energy
  deposition in the calorimeter.
\end{itemize}
The intention of this topological trigger condition is essentially to reduce the
internal background originating from the LYSO crystals at the inner layer of the
calorimeter, while at the same time ensuring the detection of good events coming from
top of the detector.

A comprehensive study of trigger efficiency in terms of reduction in background
counts in the detector is carried out, and the results are listed in Table~\ref{tab:trigeff}.
The particles used for this calculation are chosen keeping in mind different background
sources that affect the observation described in Section~\ref{sec:bkg}. Whereas the
particles and radiation from outside the detector are isotropically distributed
over different parts of the sky with respect to the detector, intrinsic radioactivity
is considered homogeneously distributed over the LYSO crystal volumes contributing in
this background component.

\begin{table}[htp]
  \centering
  \begin{tabular}{p{5.5cm} c c}
        \hline
        Background type                                                   & Basic trig. & Basic + topo. \\
                                                  & [\%]        & trig. [\%] \\
        \hline
        $\gamma$ from upper hemisphere                    & 25.1            & 29.5 \\
        $\gamma$ from lower hemisphere                    & 43.3                & 47.4 \\
        n from lower hemisphere                           & 54.9            & 61.6 \\
        e$^-$ from all directions                         & 91.0                & 91.6 \\
        e$^+$ from all directions                             & 85.7            & 86.5 \\
        p from all directions                             & 91.8                & 91.9 \\
    p from upper hemisphere                       & 98.7                & 98.9 \\
    Intrinsic radioactivity in LYSO                       & 97.8                &  99.0 \\
        \hline
  \end{tabular}
  \caption{Efficiency of the trigger conditions in reducing the background counts in the
  detector (LYSO\,+\,GAGG configuration) for different background sources.}
  \label{tab:trigeff}
\end{table}

\subsection{Effective area}
\label{ssec:effarea}

In order to estimate the detection power of the instrument, the effective area
of the detector module is calculated. To achieve this, the simulation generates
sets of parallel photons coming from random positions on a plane placed at different
directions with respect to the detector. These photons impinge on the whole detector.
To emulate a distant photon source, the simulated photons are generated from a
32 cm $\times$ 32 cm square source plane (covering the whole dome projection),
placed at a distance of 16 cm from the center of the detector but in various
directions. Normal to the source plane at its center is always directed towards the
center of the detector dome. Parallel photons are generated in a direction
perpendicular to the plane. The energy distribution of the generated photons follows
a power law with a spectral index of $-$1 (i.e., flat on the logarithmic scale),
in an energy range of 30 keV to 100 MeV. One of the key features of CE geometry is
that it provides an almost uniform response across its FoV. To study this response of
the detector, photon source planes were considered at different zenith angles at
1$^{\circ}$ apart covering the full 0--90$^{\circ}$ range, while keeping a fixed
azimuth, using the advantage of the azimuthal symmetry of the detector design.

\begin{figure}[htp]
  \centering
  \includegraphics [height=0.36\hsize]{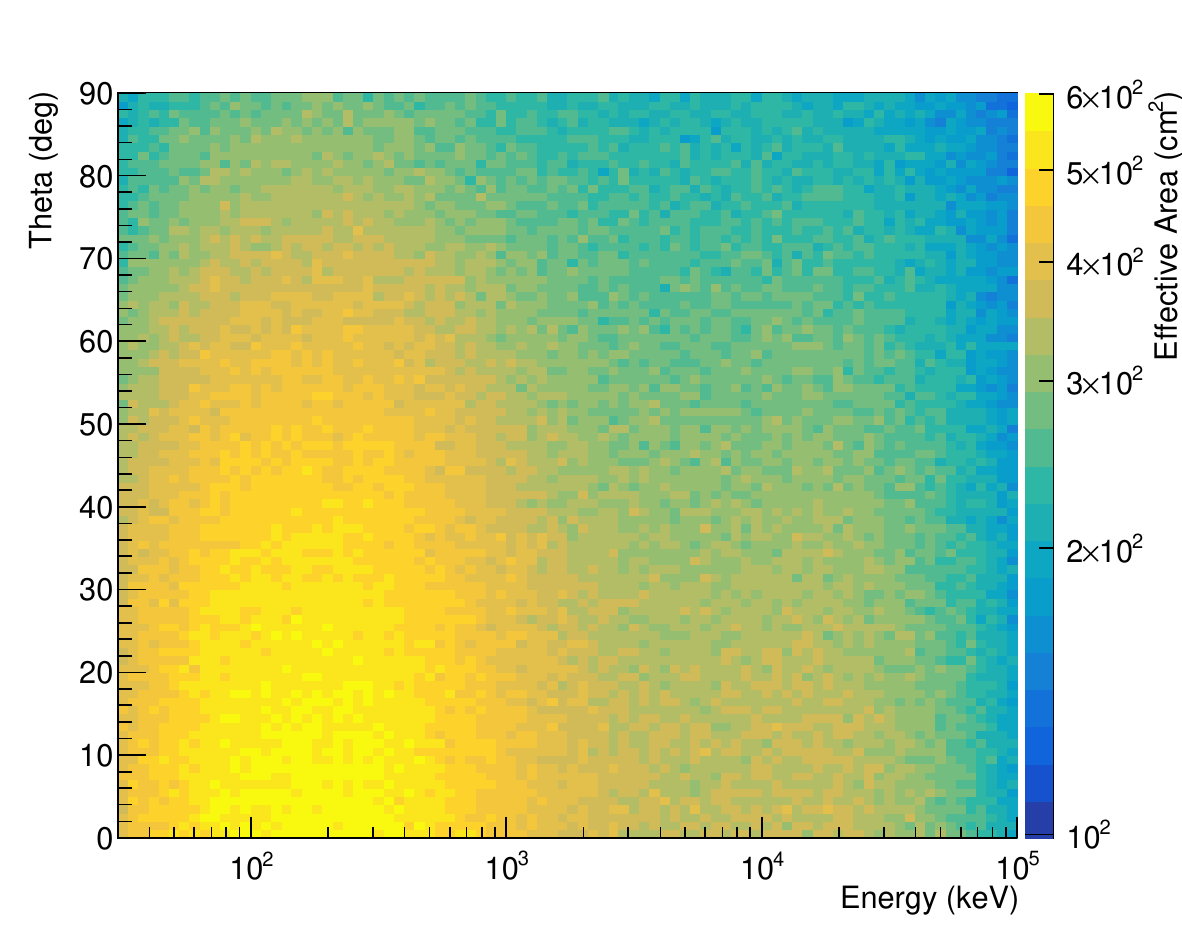}
  \includegraphics [height=0.36\hsize]{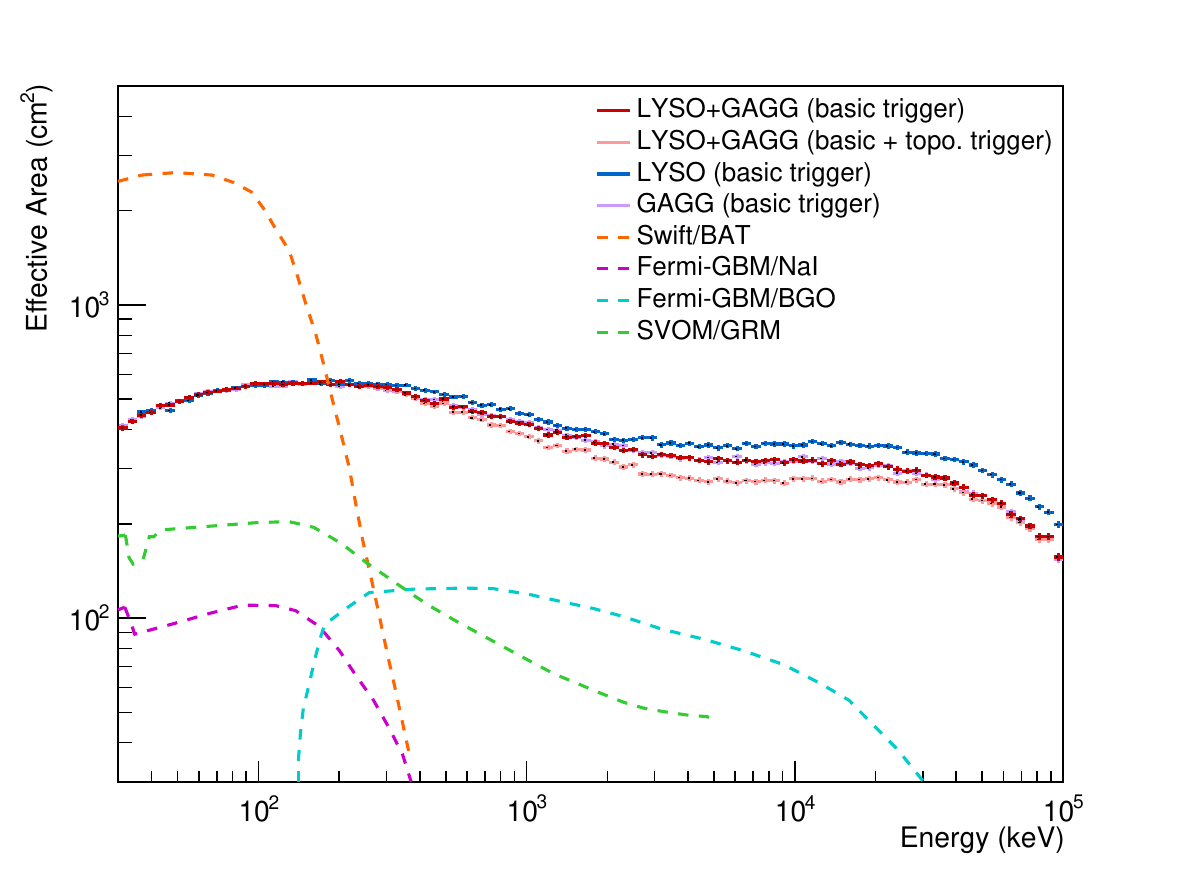}
   \caption{Left: Crystal~Eye effective area as a function of energy and zenith angles
   of the source location. Right: Crystal~Eye effective area as a function of energy
   for a source located at zenith ($\theta$ = 0$^{\circ}$). Comparative effective areas
   for different configurations (LYSO, LYSO\,+\,GAGG, GAGG) with basic trigger condition are
   shown, whereas result with basic\,+\,topological trigger condition is also plotted for
   LYSO\,+\,GAGG. The effective areas of other experiments in the similar energy range are
   also shown for comparison.}
  \label{fig:effarea}
\end{figure}

The effective area ($A_\mathrm{eff}$) is calculated in the following way:
\begin{equation}
  A_\mathrm{eff} = \frac{N_\mathrm{sel}}{N_\mathrm{sim}} \times A_\mathrm{src} \\
\end{equation}
where $N_\mathrm{sel}$ is the number of events that pass the selection cuts and
$N_\mathrm{sim}$ is the number of events generated from the source surface of the
area $A_\mathrm{src}$. The left panel of Fig.~\ref{fig:effarea} shows the detector
effective area (for the LYSO configuration with the basic trigger condition) for
different photon energies and zenith angles of the source. The detector efficiency
at different energies, manifested in the effective area plot, is the result of the
applied selection cuts convolved with the crystal absorption efficiency (see
Section~\ref{ssec:pixel}). The apparent change of the effective area for the sources
at the higher zenith angle is due to the fact that the projected geometrical area
of the detector dome changes with the zenith angle. For example, while a source
located at the zenith with respect to the detector ``sees'' a whole circular
projection of the detector dome, a source at $\theta$ = 90$^\circ$ sees only half
disc, thus reducing the effective area.

We calculated the effective area of the detector with a source at the zenith ($\theta$ =
0$^{\circ}$) for all detector configurations (LYSO, LYSO\,+\,GAGG, and GAGG) and considering
the basic trigger condition. This shows a small difference between LYSO and LYSO\,+\,GAGG
(mainly in the higher energies), as depicted in the right panel of Fig.~\ref{fig:effarea},
while LYSO\,+\,GAGG and GAGG show almost the same effective area. The situation with
the basic\,+\,topological trigger condition for the LYSO\,+\,GAGG is also shown in the
same figure, indicating that it is primarily affected in the middle energy region, where
the Compton scattering effect is most dominant. A comparison with other detectors in
similar energy range given in the same figure shows that CE provides a better effective
area by a few factors than Fermi-GBM \citep{meegan_2009}, and SVOM/GRM \citep{he_2025}.
Although SWIFT/BAT \citep{barthelmy_2005} have a better effective area than CE, it is
operative only in the lower part of the energy range covered by CE.

\section{Detector background}
\label{sec:bkg}

The CE detector modules can be used in different modes and locations in space, for example,
as a free flyer, part of a complex satellite, onboard space station, or on the Moon surface.
Here, in this study, we consider a particular situation that is intended to operate in a
circular low-Earth orbit (LEO) at an altitude of about 550 km and with an inclination of
20$^{\circ}$. So, the spacecraft is assumed to transit through relatively low-background
equatorial regions, away from the South Atlantic Anomaly (SAA) \citep{badhwar_1999} and
polar regions. In this environment, the dominant background radiation may be assumed from
the cosmic origin and secondary products from the interaction of these cosmic radiations
with the atmosphere. These include cosmic diffused $\gamma$-ray photons; albedo X-ray
and $\gamma$-ray photons from the Earth's atmosphere; albedo neutrons; trapped e$^-$
and e$^+$ in the Earth's magnetic field; and trapped protons and primary protons in the
cosmic rays.

However, apart from these external backgrounds due to the orbital radiation environment,
there can be other sources of detector background. Among them, generation of the internal
radioactivity in the heavier elements from the detector, due to activation or spallation
process by the high-energy cosmic-ray particles, can be a significant contributor. Another
contribution owing to the presence of natural radioactive isotopes in the detector materials
may also be crucial. The background from the activation process has been implicitly handled
by considering the radioactivation process in the physics-list used during the simulation
of the external radiation interaction. In contrast, the radiation background from the natural
isotopes has been explicitly calculated in this exercise considering the information from
the observed activity in the material. This is particularly important for the LYSO
crystals, which exhibit significant radioactivity in a limited energy range.

\subsection{Background from the orbital radiation environment}
\label{ssec:extbkg}

The background counts in the astronomical radiation detectors obviously depend on the
operating radiation environment. The effect also depends on the distribution of materials
in the detector and its surroundings through the generation of secondary radiation
and particles. The distribution of different background components at the LEO has been
discussed in several works such as \cite{ajello_2008}, \cite{mizuno_2004}, \cite{sarkar_2010}
(and the references therein). For current purposes, we use the calculations given by
\cite{cumani_2019} to predict the differential flux of various particle and radiation
components at the operational orbit of CE while using a moderate solar modulation potential
(650 MV). Although the models used to describe the individual background components can be
found in more detail in \cite{cumani_2019}, here we briefly discuss them in the context of
their use in this current simulation.

One of the dominant contributors to the background of photon detectors in the energy
range of our interest is the diffused cosmic photons. This isotropic background is
believed to be the combination of the integrated emission of active galactic nuclei
and other unresolved extragalactic sources \citep{ajello_2008}. High-energy cosmic-ray
interactions with Earth's atmospheric nuclei produce hadronic and EM cascades
including muons and other hadrons. Whereas the production of $\gamma$ rays above 50 MeV
is associated with the decay of mesons, at lower energies it can be accounted for the
bremmstrahlung radiation from secondary electrons. Although an asymmetry is expected
due to the effect of Earth's magnetic field in the charged component of the shower
(mainly protons), it has been shown that the effect is negligible for the keV to low-MeV
regime \citep{abdo_2009}. Therefore, albedo photon emission can be considered isotropic
across the surface of the Earth, which affects the detector from the bottom. Albedo neutrons
are also generated by the interaction of cosmic rays with the atmosphere, and can reach
at LEO to interact with the detector materials, giving rise to background counts. Also, in
this case, we consider an isotropic distribution of the flux, but from the lower part of
the detector only, similar to the albedo photons.

The direct interaction of cosmic-ray protons with the instrument can generate a signal
after the material de-excitation or induced radioactivity. CE orbit is relatively low
in altitude and inclination, so the magnetic field shielding largely protects the
instrument from low-energy protons ($\lesssim$ GeV). The geomagnetic cutoff rigidity for
primary protons in the operating orbit of CE ($\gtrapprox$ 10 GV) is well beyond the upper
limit of its energy range. So, the expected contribution to the detector background from this
component is low. However, there may be some contribution through the secondary generation
and activation process in the detector material. The interaction of high-energy cosmic rays
with the atmosphere can also produce secondary charged particles (protons and e$^-$/e$^+$)
that constitute an additional background in the operation energy range of CE. The trapped
protons in the geomagnetic field from the decay of albedo neutrons can also add to this
component along with the charged particles from solar wind. There may be some anisotropy
in the secondary charged particle flux due to the distribution of the interaction probability
of primary particles in the atmosphere and magnetic field consequences like the East-West
effect. But, for the sake of simplicity and considering the almost uniform response of the
detector in a wide FoV, which averages out the anisotropy to some extent, we consider an
isotropic flux distribution of the secondary charged particles from all the directions.

\begin{figure}[htp]
  \includegraphics [width=0.49\hsize]{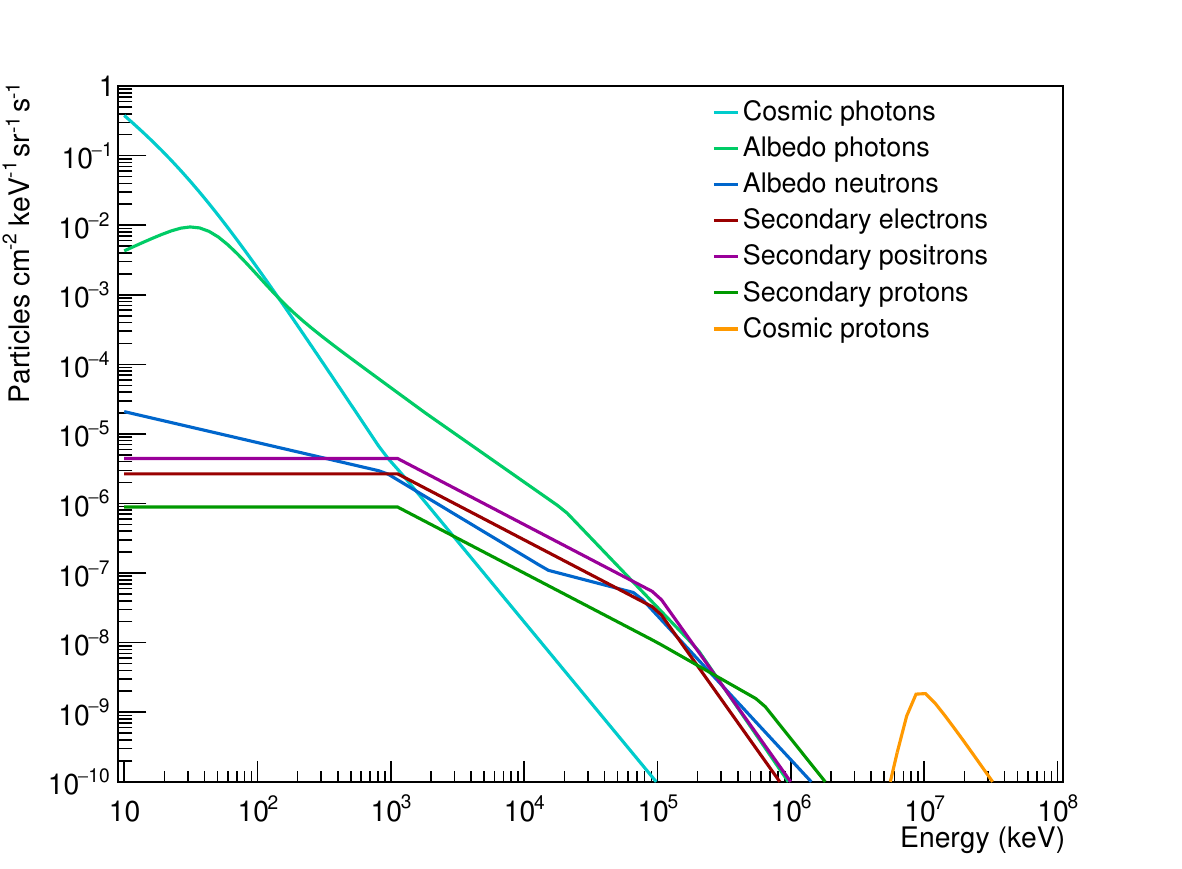}
  \includegraphics [width=0.49\hsize]{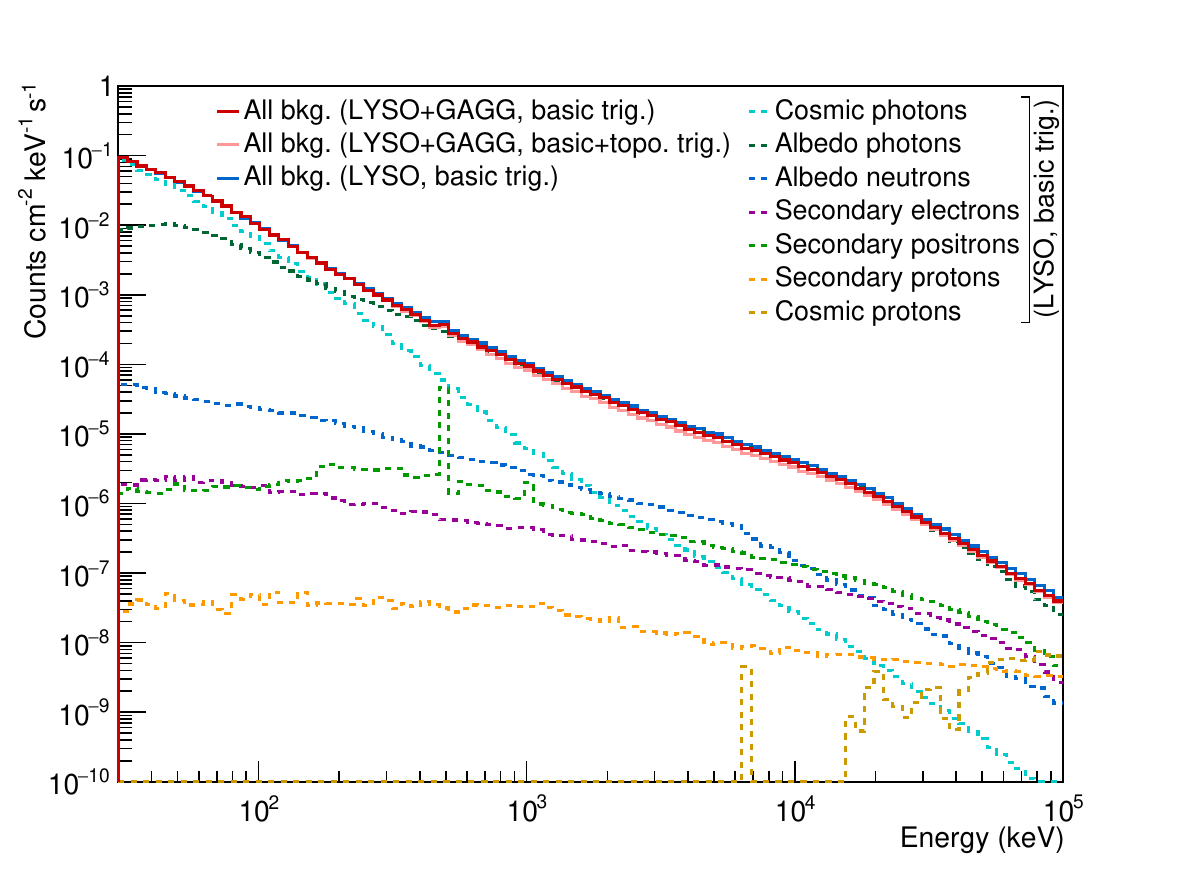}
  \caption{Left: Differential flux models of the particles and radiation components
  dominating the CE orbital radiation environment. Right: Estimated background fluxes in
  CE due to different orbital radiation components. Total background is shown  with basic
  trigger condition for both LYSO and LYSO\,+\,GAGG configurations while contributions
  from individual components are for LYSO. The case of basic\,+\,topological trigger
  condition is also calculated for LYSO\,+\,GAGG configuration. Background with only
  GAGG is not shown here as it would be overlapped by the LYSO\,+\,GAGG background curve.}
  \label{fig:extbkg}
\end{figure}

The differential spectra of all potential background components at LEO near the equatorial
region that is supposed to host the CE experiment are plotted in the left panel of
Fig.~\ref{fig:extbkg}, in a wider range of energy than the upper limit of CE.

\subsection{Simulation of the orbital background}
\label{ssec:orbbkgsimu}

To properly estimate the detector response to radiation and particle backgrounds,
detailed simulations were carried out considering the interactions of incident
particles with the overall detector construction described in Section~\ref{sec:detDesc}
and \ref{ssec:simu}. We calculated the energy depositions in every crystal and veto
layer of the detector for the different background particles. Depending on the interaction
scenario of the particles and their effective contribution in the operating energy range
of the detector, we chose different energy ranges for the incident particles to increase
the efficiency of the simulation procedure. For example, we considered incident photons
in the energy range of 30 keV to 1 GeV; neutrons in 30 keV to 10 GeV; secondary protons
in 10 MeV to 1 GeV; cosmic protons in 4 GeV to 100 GeV; e$^-$ and e$^+$ in 30 keV to 1 GeV.
However, in case of the secondary charged particles, since the calculation of \cite{cumani_2019}
only covers the energy range stopping at 1 MeV, we extrapolated these fluxes up to 30 keV
with constant fluxes corresponding to the particles at 1 MeV (see Fig.~\ref{fig:extbkg}).
This consideration is reasonable, as most of the incident particles in this low-energy
range are suppressed by the veto layers, leaving a negligible effect in the detector
background for a moderate change in the extrapolated flux.

From the simulation point of view it is also important to watch out the simulation
efficiency, i.e., how many particles generated from the source surface are going to actually
traverse through the detector structure and contribute in the event detection statistics.
To ensure a better simulation efficiency for the background estimation, we considered
a concentric hemispherical source surface of 16 cm radius around the detector dome,
to cover the upper part of the detector. In contrast, the particles coming from the
bottom of the detector are generated from a circular plane of the same radius placed
just below the bottom ACD. Particles or photons are randomly produced from these surfaces,
with their direction randomly distributed according to the cosine law in the 0--90$^{\circ}$
angular range with respect to the surface normal at the generation point. This particular
scheme for the source-surface arrangement minimizes ``leakage'' of the primary particle
tracks without passing through the detector structure. Whether the randomization in the
direction ensures the isotropic nature of the background flux distribution at the detector.
Depending on the expected flux distribution of different background components, we simulated
cosmic photons and primary protons from the upper hemisphere; albedo photons and neutrons
both from the upper hemisphere and lower circular plane but considering only those particles
moving in the upward direction; and secondary charged particles both from the upper hemisphere
and lower circular planes going in all directions. In most of the cases, we simulated 10$^6$
particles for each run (except for primary protons, where we simulated 10$^5$ particles
to save simulation time, since the contribution from this component is not so substantial).

Events are originally sampled from an energy distribution following a power law with a
spectral index of $-$1, corresponding to a flat distribution on the logarithmic scale. After
applying trigger selection and threshold cuts, as mentioned in Section~\ref{ssec:trig},
a weighting procedure is performed to get the actual background contributions for
each of the components according to their input flux distribution. Although the incident
energy ranges are different for different particles, the deposited energy range is
always fixed from 30 keV to 100 MeV. The upper energy limit is considered a little bit higher
than the preliminary proposed limit to see the detection effect in the higher energy, which
will help to optimize the energy range of the experiment in the future considering other
constraints by readout electronics, data budget, etc. In the analysis, both the incident
and deposited energy ranges are divided into 100 bins evenly spaced in logarithm of energy
values. The weighted count rates in the deposited energy bins are calculated as:
\begin{equation}
  dN_j = \sum_i \int_S \int_\Omega \int_{E_i}^{E_{i+1}} \Phi(E_i)\,dE_i\,d\Omega\,dS\,
  \frac{N_{ij,\,\mathrm{dep}}}{N_{i,\,\mathrm{gen}}}.
  \label{eq:weight}
\end{equation}
Here, $i$ and $j$ are the energy bin indices over the incident and deposited energy,
respectively. $N_{i,\,\mathrm{gen}}$ is the number of particles generated in each incident
energy bin and $N_{ij,\,\mathrm{dep}}$ is the number of events with energy deposition in the
$j^{th}$ bin due to all incident events in the $i^{th}$ bin. $\Phi(E_i)$ represents the
incident flux spectrum in units of particles\,cm$^{-2}$\,sr$^{-1}$\,keV$^{-1}$\,s$^{-1}$
for different particles as described in Section~\ref{ssec:extbkg}. The integration of
particle flux is done on the energy ($E$), the area of the source surface ($S$),
and the solid angle ($\Omega$) of the randomized direction of the incident particles. The
final energy deposition spectra is expressed in counts\,cm$^{-2}$\,keV$^{-1}$\,s$^{-1}$,
and is obtained by dividing $dN_j$ by the geometrical area of the detector and the width
of the deposited energy bins.

The partial and total contributions of the detector background due to different components
of the external radiation and particles are shown in the right panel of Fig.~\ref{fig:extbkg}.
Total background counts are calculated for all the configurations with the basic trigger
condition. The background for the LYSO\,+\,GAGG configuration using the basic\,+\,topological
trigger condition is also shown in the same plot. However, the result for the GAGG only
configuration is not shown in the plot since it would be overshadowed by the LYSO\,+\,GAGG
background plot, which is also evident from the effective area plot in Fig.~\ref{fig:effarea}.
It is clear that the dominant background component comes from $\gamma$ radiation, as expected.
Below about 200 keV the cosmic diffused photon is, by far, the predominant component. At higher
energies, the albedo $\gamma$ component becomes prevalent. Particle backgrounds are more
significant only at higher energies. The effect from the primary cosmic-ray protons is only
noticeable at the highest part of the energy range, therefore carries not much significance
for CE background. It is also apparent that the external background is quite similar for
all the detector configurations, since the densities and interaction probabilities of particles
and radiation in the LYSO and GAGG are comparable. The effect of topological trigger
is marginally visible in the Compton-effect dominated energy region, as is also visible in
the effective area calculation. The total integrated background rates in the energy range
of 30 keV to 100 MeV, for different detector configurations and trigger conditions, are
given in Table~\ref{tab:bkg}.

\subsection{Intrinsic radioactive background}
\label{ssec:intbkg}

Despite the several advantages of using LYSO as a scintillator material, there is one major
challenge using this crystal roughly in the 100 keV to 2 MeV energy range due to the presence
of intrinsic radioactivity. LYSO crystals are naturally contaminated by the presence of
radioactive isotope $^{176}$Lu with an approximate activity of 40 Bq\,g$^{-1}$.
$^{176}$Lu has a half-life of $\sim$ 10$^9$ years and undergoes $\beta^-$ decay
with a maximum energy of 593 keV. Subsequently, three prompt $\gamma$ rays are emitted
in the decay process with their corresponding probability, with approximate energies of
88, 202, and 307 keV. These $\gamma$ rays and the electron from the $\beta^-$ decay
deposit their energies in the detector crystals according to their interaction probabilities,
giving rise to the intrinsic background.

We simulated this intrinsic background in the detector considering random $^{176}$Lu
isotopes uniformly distributed in all of the LYSO crystals (10$^6$ events were generated for
the calculation) and let them decay. The required time interval for this number of decays
was calculated using the activity rate and the total amount of LYSO material. Each event was
assigned with a time stamp in this calculated time interval. The random coincidence between
independent events that undergo $\beta^-$ decay inside the decay-time window of the
event signal in the calorimeter readout system (taken to be 90 ns) was also considered
for the calculation.

\begin{figure}[htp]
  \centering
  \includegraphics [width=0.5\hsize]{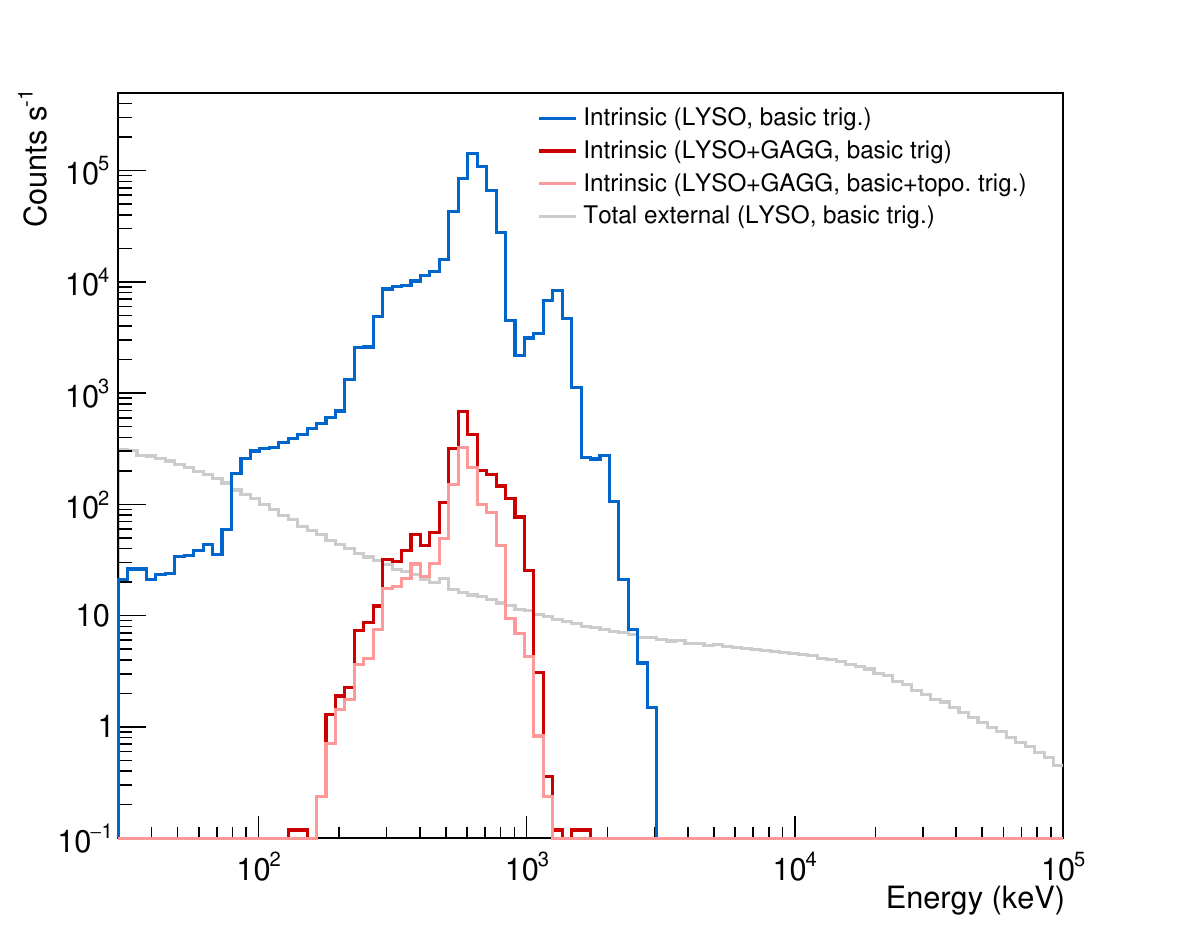}
  \caption{Estimated intrinsic background in the detector due to the natural radioactive
  isotopes in the LYSO crystals. For LYSO\,+\,GAGG configuration both the results with
  basic and basic\,+\,topological trigger conditions are shown, while for LYSO only
  configuration basic trigger condition is used. The total contribution from the
  orbital background in the case of LYSO with basic trigger is also shown for
  comparison.}
  \label{fig:intbkg}
\end{figure}

The count rate spectrum of the intrinsic background is shown in Fig.~\ref{fig:intbkg}
along with the total external background (in LYSO configuration). The contribution
substantially exceeds the orbital background in the energy region with a peak near
about 600 keV (combined energy of the three $\gamma$-ray emissions and $\beta^-$
decay). A higher background contribution from the LYSO only configuration is obvious,
owing to the higher amount of radioactive material. The peak near 1200 keV and
background counts beyond are due to coincident events in the time window of the
signal-pulse decay, added with the electron energy from the $\beta^-$ decay. The total
background count rates over the whole energy range for different detector and trigger
configurations are given in Table~\ref{tab:bkg}. The presence of this background
component gives rise to a serious concern for the detector sensitivity and is
discussed further in Section~\ref{sec:sensitivity}. On the other hand, this peak
can in principle be used as the onboard calibrator for the detector.

\begin{table}[htp]
  \centering
  \begin{tabular}{p{7cm} c c}
        \hline
        Configuration                         & External & Intrinsic \\
                                              & [kHz]    & [kHz] \\
        \hline
        LYSO, basic trig.                     & 4.479    & 599.194 \\
        LYSO\,+\,GAGG, basic trig.            & 4.419    & 2.570 \\
        LYSO\,+\,GAGG, basic\,+\,topo. trig.  & 4.369    & 1.143 \\
        \hline
  \end{tabular}
  \caption{External and intrinsic background count rates in the detector for
  different crystal materials and trigger conditions.}
  \label{tab:bkg}
\end{table}

\subsection{Background count distribution in pixels}
\label{ssec:pixbkg}

\begin{figure}[htp]
  \centering
  \includegraphics [width=0.5\hsize]{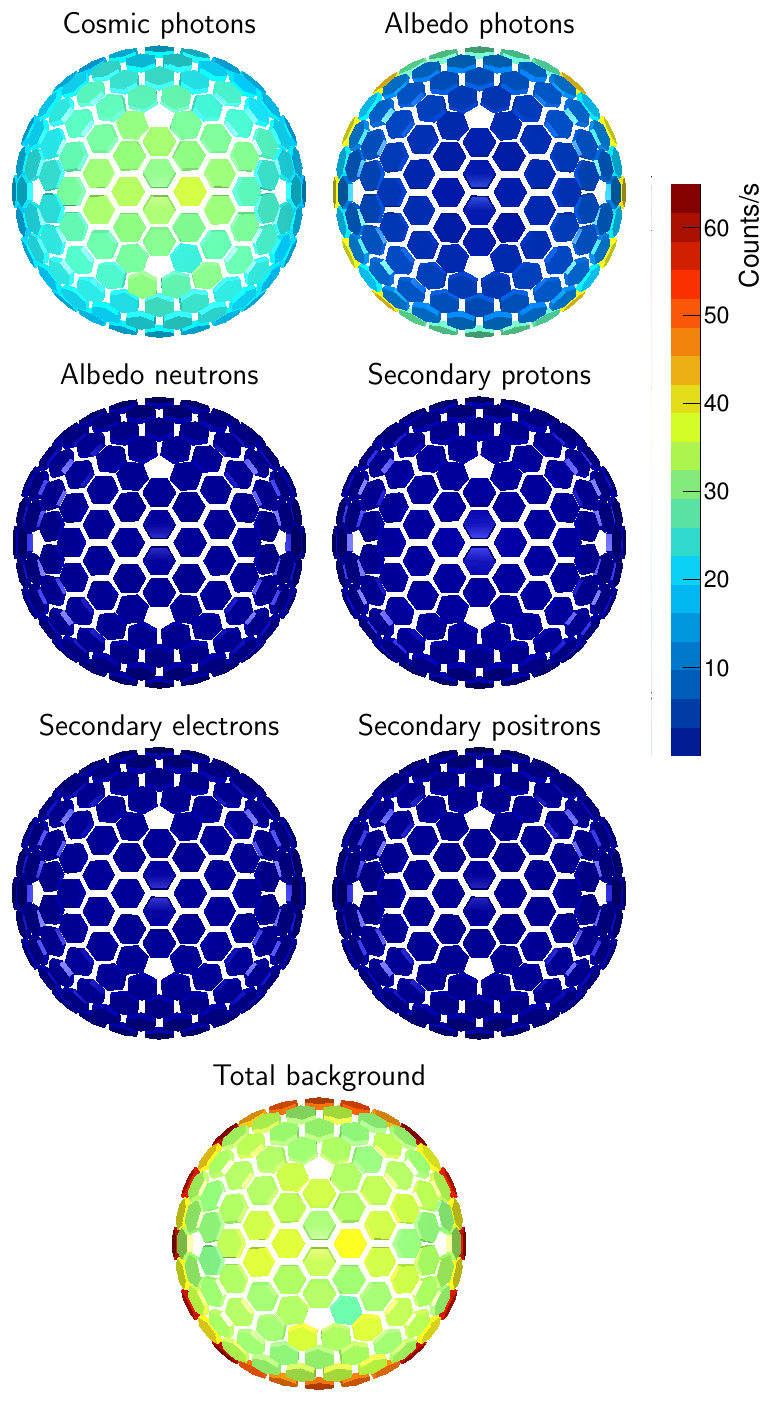}
  \caption{Distribution of integral count rates of weighted count rates in the
  pixels of CE detector (LYSO configuration) due to the orbital background components.
  Total contribution for all the components is also shown.}
  \label{fig:pixbkg}
\end{figure}

In order to have an idea of the spatial distribution of different components
of the orbital background in the detector, we calculated the weighted count
rates in different pixels (top and bottom crystals) for the individual background
components. Figure~\ref{fig:pixbkg} shows the integral count rates in the detector
pixels for a LYSO only configuration with the basic trigger condition for different
background components, as well as the total contribution for all external backgrounds.
We do not show the background for cosmic-ray protons as the contribution is
negligible compared to others.

The count distributions are apparently consistent with the expected response for
isotropic backgrounds. For cosmic backgrounds coming from outer space, the
dome-like geometry of the CE detector results in a more or less uniform count
distribution over the detector pixels. However, some events hitting the lower part
of the detector dome may be cut off due to the containment condition in the calorimeter
(i.e., shower is not contained in the detector crystals). The albedo backgrounds
are more concentrated among the pixels near the bottom of the detector dome. This
is because the lower structure blocks most of the incoming particles, and only those
hitting the edge of the detector deposit their energy in the pixels. However, the
secondaries produced in the lower structure would first interact in the pixels at
the bottom of the dome and deposit their energy. The total contribution from the
external background is shown in the bottom panel of Fig.~\ref{fig:pixbkg}. In
addition to the external backgrounds, intrinsic radioactivity will also contribute
to the detector background with a uniform distribution over all pixels, as shown
in Fig.~\ref{fig:pixgrb}.

\section{Sensitivity and source response}
\label{sec:sensitivity}

The sensitivity of an instrument establishes the minimum flux necessary for significant
detection, quantifying the capability to observe different phenomena on top of the
background level. Although the detector has been designed and optimized mainly for the
detection of short-term transient sources in the sky, nevertheless it also has some
capability to observe bright persistent sources in its FoV. Thus, the detector can also be
used to study the interesting activities of these sources in a less-explored energy range.
The sensitivity for persistent sources can be calculated considering these point-like
sources over a relatively larger period of time (typically of the order of year). This
requires precise background estimation, ephemeral information of the target source and
other sources in the detector FoV, and also using some special techniques such as Earth
or Moon occultation \citep{wilson_2012}. Other technical aspects like the stability of
detector gain during long-term observation for these kinds of source should also be
taken into consideration. While a detailed calculation of the detector sensitivity for
the persistent sources will be performed and reported in the future, here we present
sensitivity of the detector for transient sources.

The sensitivity of the detector is calculated to evaluate the detection capabilities
when considering transient sources that are short-duration transient phenomena with
a timescale typically of the order of seconds, such as GRBs. To calculate the transient
sensitivity of the detector, an approach similar to the one used in \cite{martinez_2022}
is adopted. The signal-to-noise ratio (S/N) for transients is defined as
\begin{equation}
  N_\sigma = \frac{N_\mathrm{S}}{\sqrt{N_\mathrm{S} + N_\mathrm{B}}},
  \label{eq:snratio}
\end{equation}
where $N_\mathrm{S}$ corresponds to the number of source events and $N_\mathrm{B}$ the
number of background events integrated over the observation or exposure time interval
$\Delta T$. The required source flux can be computed to meet a given $N_\sigma$ detection
threshold. Here, the transient sensitivity of the detector is calculated for two nominal
exposure times: $\Delta T$ = 2 s and $\Delta T$ = 8 s, which are between the duration of
short and long GRBs. (These choices are also, to some extent, influenced by the preliminary
consideration of the event triggering and accumulation in the onboard data processing
procedure.) The total background is calculated from all components of the simulated
background as described in Section~\ref{sec:bkg}, including the intrinsic radioactivity
of the LYSO crystals.

To calculate the signal events, we considered typical GRB spectra. The GRB spectral
models can be defined by: Band function \citep{band_1993}, exponentially attenuated
power-law function (hereafter referred as ``Comptonized''), a single power-law function,
and other spectral models \citep{poolakkil_2021}. However, in this work, we consider either
Band or Comptonized functional forms for the GRB spectral representation. The Band GRB
function has the form
\begin{equation}
  \Phi_\mathrm{band}(E) = A 
  \begin{cases}
    \left(\frac{E}{100\,\mathrm{keV}}\right)^{\alpha}\,exp{\left[- \frac{(\alpha\,+\,2)\,E}
    {E_\mathrm{peak}}\right]} & \\
    \hspace{3cm} \mathrm{for}\,\,E \leq \frac{(\alpha\,-\,\beta)\,E_\mathrm{peak}}{\alpha\,+\,2} & \\[5pt]
    \left(\frac{E}{100\,\mathrm{keV}}\right)^{\beta}\,\left[\frac{(\alpha\,-\,\beta)\,E_\mathrm{peak}}
    {(\alpha\,+\,2)\,100\,\mathrm{keV}}\right]^{(\alpha\,-\,\beta)}\,exp(\beta - \alpha) & \\ 
    \hspace{3cm} \mathrm{otherwise,} & \\
  \end{cases}
  \label{eq:band}
\end{equation}
where $A$ is the amplitude, $\alpha$ and $\beta$ are the low- and high-energy power-law
indices, respectively, and $E_\mathrm{peak}$ is the characteristic peak energy. The
Comptonized function can be described only by the low-energy part of the Band function, i.e.,
\begin{equation}
  \Phi_\mathrm{comp}(E) = A\,\left(\frac{E}{100\,\mathrm{keV}}\right)^{\alpha}\,
  exp{\left[-\frac{(\alpha + 2)\,E}{E_\mathrm{peak}}\right].}
  \label{eq:comp}
\end{equation}
We simulated parallel photons in the detector from a square surface (32 $\times$
32 cm$^2$), placed at the zenith of the detector ($\theta$ = 0$^\circ$), while the
energy is distributed by a flat spectrum on logarithmic scale. Then we calculated
the energy response in the detector and weighted the deposited spectrum by the incident
GRB spectrum. We used the Fermi-GBM GRB spectral catalog \citep{poolakkil_2021}
to obtain spectral information for 2300 GRBs available in the catalog, considering
the Band and the Comptonized spectral models.

\begin{figure}[htp]
  \centering
  \includegraphics [width=0.49\hsize]{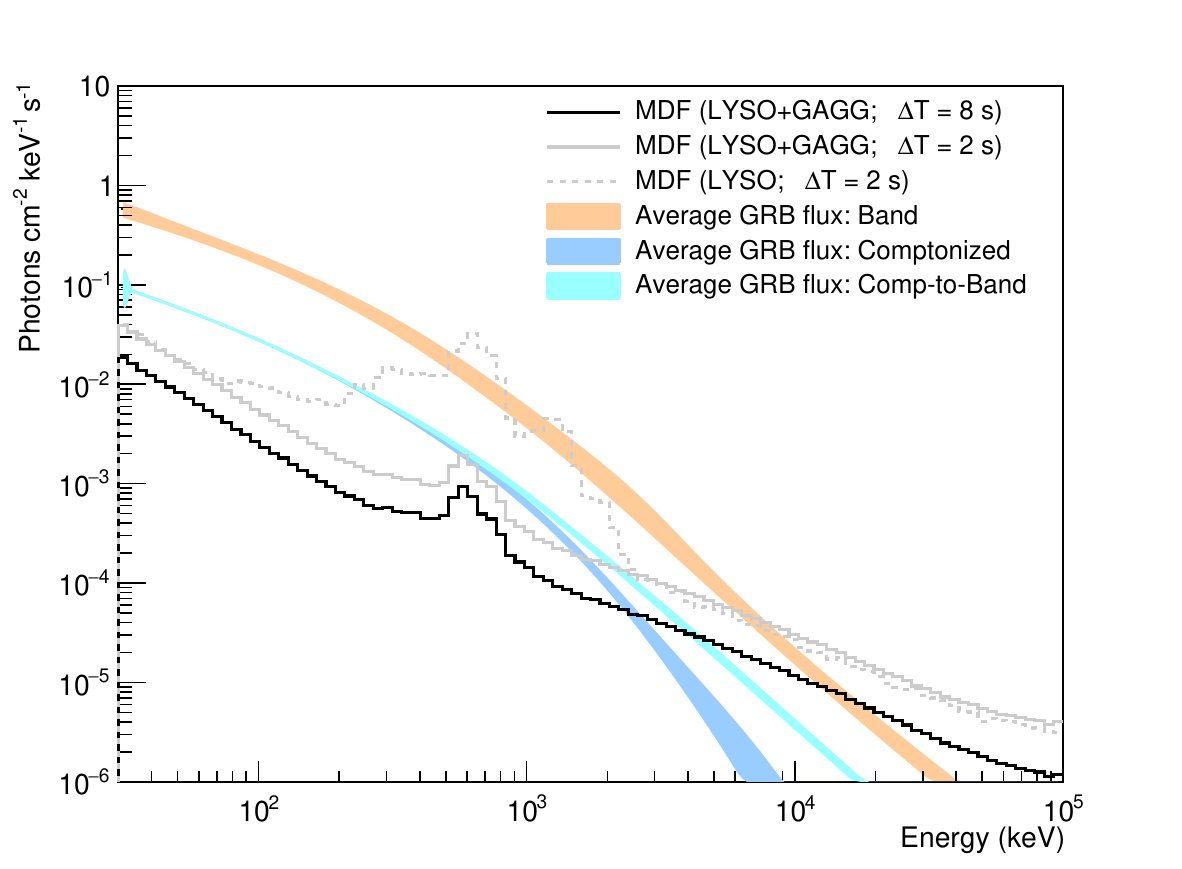}
  \includegraphics [width=0.49\hsize]{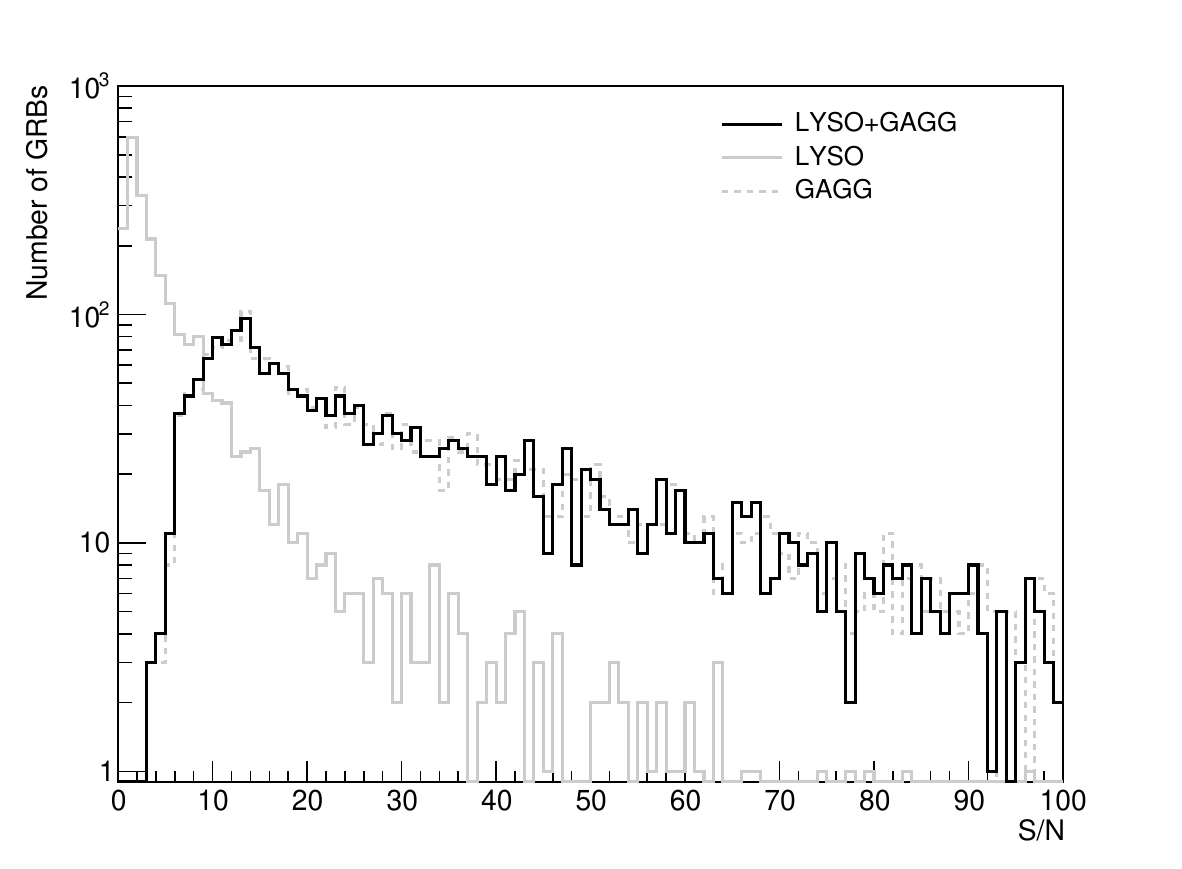}
  \caption{Left: Minimum detection flux of the detector for GRB detection. Different
  MDF of the detector is shown considering 2 s of exposure time for both LYSO with
  basic trigger and LYSO\,+\,GAGG with basic\,+\,topological trigger configurations of the
  detector. The MDF for 8 s exposure for LYSO\,+\,GAGG is also shown. The average GRB
  spectra calculated from the Fermi-GBM catalog shown with a band of 1$\sigma$
  standard error for best-fit Band function, Comptonized function, and Comptonized
  functions modified to Band (Comp-to-Band) are also shown for visual comparison.
  The calculation is done considering the source located at the zenith ($\theta$
  = 0$^{\circ}$). The results for GAGG only configuration are not explicitly shown here,
  because they will be similar to LYSO\,+\,GAGG but without the peak due to intrinsic
  background. Right: Signal-to-noise ratios for all the GRBs from the Fermi-GBM
  catalog for all configurations of the detector. The S/N values are calculated for
  $\Delta T$ = 2 s and considering Comptonized model of all the GRB spectra.}
  \label{fig:snr}
\end{figure}

The minimum detectable flux (MDF) for the detector as a function of energy is calculated
by solving Eq.~\ref{eq:snratio} for $N_\mathrm{S}$ in each energy bin which is
\begin{equation}
  N_\mathrm{S}(E) = 0.5\,N_\sigma^2 \left(1 + \sqrt{1 + \frac{4\,N_\mathrm{B}(E)}{N_\sigma^2}}\right).
  \label{eq:mdl}
\end{equation}
With $\Phi_\mathrm{min,\,trn}(E)$ as the minimum incident flux required for a transient
detection, $N_\mathrm{S}(E)$ $\approx$ $\Phi_\mathrm{min,\,trn}(E) \cdot A_\mathrm{eff}(E)
\cdot {\Delta}E \cdot \Delta T$, where $A_\mathrm{eff}(E)$ is the effective area of the
detector in cm$^2$ and ${\Delta}E$ is the energy bin width.\footnote{This is true for
events with full energy deposition in the calorimeter, i.e., contained events, otherwise
energy response matrix should be used instead of $A_\mathrm{eff}(E)$ for precise
conversion.} $N_\mathrm{B}(E)$ = $B(E) \cdot \Delta T$, where $B(E)$ is the integrated
background count rate in each energy bin (i.e., total background from orbital and
intrinsic sources shown in Fig.~\ref{fig:intbkg}). Thus, the MDF can be obtained as
\begin{equation}
  \Phi_\mathrm{min,\,trn}(E) = 0.5\,\frac{N_\sigma^2}{A_\mathrm{eff}(E)\,\Delta E\,\Delta T} 
  \left(1 + \sqrt{1 + \frac{4\,B(E)\,\Delta T}{N_\sigma^2}}\right).
  \label{eq:mdltran}
\end{equation}
The MDF is calculated for both LYSO and LYSO\,+\,GAGG detector configurations, with
a nominal $N_\sigma$ = 3, giving 99.85\% confidence level (C.L.). These are shown in the
left panel of Fig.~\ref{fig:snr} in units of photons\,cm$^{-2}$\,keV$^{-1}$\,s$^{-1}$.
The MDF value for the GAGG only configuration is almost similar to the LYSO\,+\,GAGG
configuration, without the peak near 600 keV due to intrinsic background, hence not
shown in the figure. For the LYSO only configuration we considered the basic trigger
condition as has been done before, but for the LYSO\,+\,GAGG we only show the
basic\,+\,topological trigger condition, as this gives slightly better sensitivity.
Hereafter we continue to use these two sets of combinations for detector configuration
and trigger condition. Here, the detector sensitivity in terms of MDF is calculated
considering a short-duration GRB with exposure time $\Delta T$ = 2 s for both detector
configurations, while for LYSO\,+\,GAGG configuration the MDF has also been calculated
for a relatively longer duration of the transient source with exposure time $\Delta T$
= 8 s, and shown in the same plot for comparison. The average GRB spectra calculated
from the spectral parameters given in the Fermi-GBM catalog for the GRB candidates, best
fitted by Band and Comptonized functions, are also shown on the same plot. Although the
Band function is usually used for a better representation of the GRB spectral form,
the Comptonized function gives a more restrictive fit to the GRB flux, particularly
showing a lower flux at the high-energy region. So for the GRB spectra best fitted by
the Comptonized function, we additionally converted them into the Band function by adding
a high-energy power-law index with a random value between $-$2.3 and $-$2.5. The resulting
average spectrum is also shown in the same plot (left panel of Fig.~\ref{fig:snr}).
All these average GRB spectra are shown by a band of 1$\sigma$ standard error.

The S/N for all GRBs detected by Fermi-GBM is calculated in the energy range of
30 keV to 100 MeV for the three detector configurations, and the results are shown
in the right panel of Fig.~\ref{fig:snr}. In this calculation, we considered a relatively
restrictive representation of the GRB spectra using the Comptonized model for all the 2300
GRBs in the Fermi-GBM catalog. It is evident from this plot that all the GRBs given
in the catalog are detectable in the LYSO\,+\,GAGG configuration with more than 3$\sigma$
C.L., but for the LYSO only configuration the situation is constrained due to the presence
of the higher intrinsic background. However, the GAGG only configuration improves the
situation from the LYSO\,+\,GAGG, but only marginally.

\begin{figure}
  \centering
  \includegraphics [width=0.5\hsize]{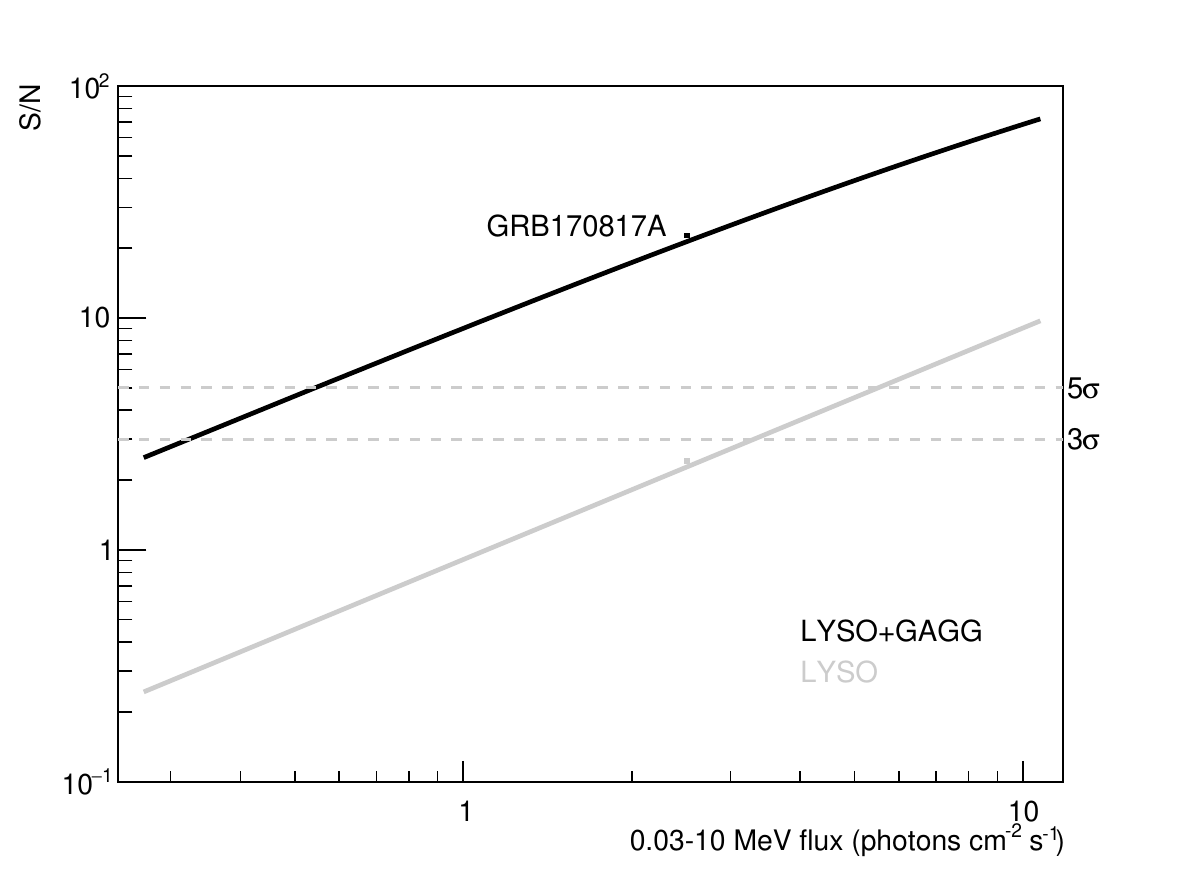}
  \caption{Signal-to-noise ratio as a function of the integrated GRB flux for a
  Comptonized spectral model with fiducial parameters and located at the zenith
  ($\theta$ = 0$^{\circ}$) for LYSO\,+\,GAGG and LYSO configurations of the detector.
  The S/N is also calculated for the source GRB170817A and shown for reference. The
  S/N values for 3$\sigma$ and 5$\sigma$ levels are marked by gray dashed lines for
  viewing guidance.}
  \label{fig:fluxsnr}
\end{figure}

In order to have an estimation of flux dependence of the GRB detection in terms of
S/N, we considered a fiducial GRB spectral model. This fiducial model is calculated
by fitting the average of all the GRB (Comptonized) spectra from the Fermi-GBM catalog
(giving $E_\mathrm{peak}$ = 756.4 keV, $\alpha$ = $-$1.07, $A$ = 0.026
photons\,cm$^{-2}$\,keV$^{-1}$\,s$^{-1}$). Keeping the spectral shape fixed and only
varying the amplitude value, we calculate the integrated flux over the 30 keV -- 10 MeV
energy range and the corresponding S/N (in the 30 keV -- 100 MeV energy range and for
2 s exposure time). Figure~\ref{fig:fluxsnr} shows the integrated flux vs. S/N plots
for the LYSO and LYSO\,+\,GAGG configurations for a source located at the zenith.
The corresponding values calculated for a particular GRB source GRB170817A (considering its
Comptonized model parameters given in the Fermi-GBM catalog and located at the zenith)
are also marked on the same plot for reference. GRB170817A is associated with the first
(and so far only combined) detection of a gravitational wave event (GW170817) with an
EM counterpart \citep{abbott_2017, abbott_2017_b, goldstein_2017, savchenko_2017}.
Again, the results of the GAGG only configuration are not shown here, as this gives a
marginal improvement over LYSO\,+\,GAGG, as shown in Fig.~\ref{fig:snr}.

\begin{figure}[htp]
  \centering
  \includegraphics [width=0.5\hsize]{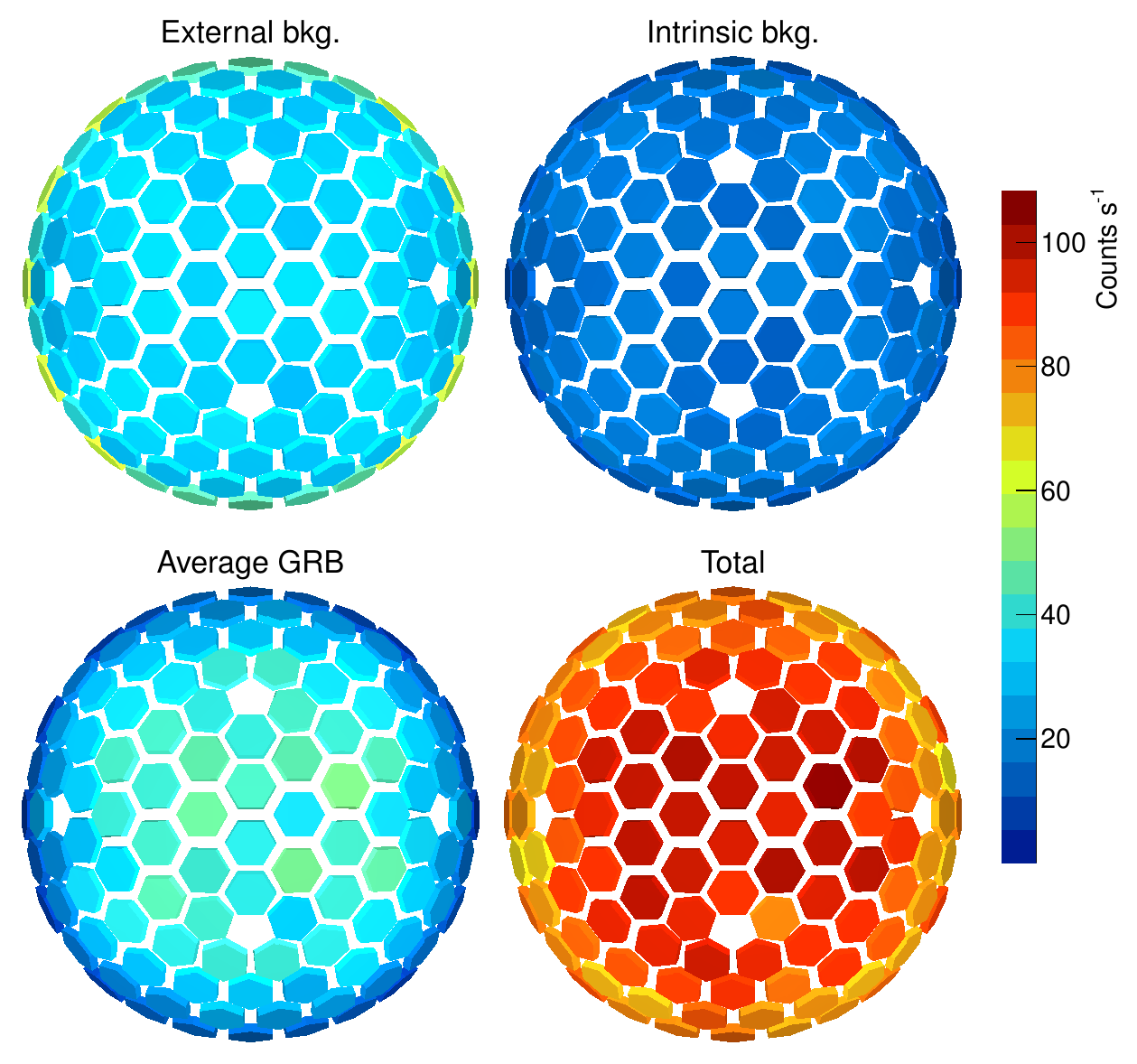}
  \caption{Background and source count rate distribution in the CE detector pixels.
  Pixels count rates for total external background, intrinsic background, and an
  average GRB source at 0$^{\circ}$ zenith angle is shown along with the total
  (source plus background) contribution.}
  \label{fig:pixgrb}
\end{figure}

To visualize the pixel distribution of a transient event in the detector with
respect to the detector background, Fig.~\ref{fig:pixgrb} shows the count rates
across the detector pixels for both external and intrinsic background sources
along with those for an average GRB source. This result is shown for the
LYSO\,+\,GAGG configuration with the basic\,+\,topological trigger condition.

\section{Real-time source localization}
\label{sec:srcloc}

One main feature of the CE detector is its capability to locate potential transient
flares in the sky and autonomously give prompt alerts to networks such as: General
Coordinates Network (GCN).\footnote{https://gcn.nasa.gov.} The localization algorithm
for transient sources is based on the image (or histogram of pixel ID vs. counts) comparison
technique. Transient sources in different directions in the sky generate signature
count distributions in the detector pixels (``pixel map''), which are similar to an
image. To identify the direction of a transient source (sample), the localization
algorithm compares the pixel map for this source with the pre-calculated (using simulation)
set of pixel maps for sources (templates) located in different directions with respect to
the detector.

The pixel maps corresponding to the sample and template sources are compared using the
Kolmogorov-Smirnov (KS) test, which gives the matching probability of the two maps.
However, the precision of the reconstructed direction of the sample source depends
on the distribution and density of the template locations, but up to certain extent
-- limited by the granularity of the detector pixels. The higher density of templates
in the angular space will give a better prediction of the reconstructed angle, but
at the expense of time and memory required for the calculation. To study this effect,
we used two sets of templates almost isotropically distributed over the sky hemisphere,
one set with approximately 5$^{\circ}$ angular separation (total 742 template sources
to cover the sky) and another set with 2$^{\circ}$ separation (4980 sources).

\begin{figure}[htp]
  \centering
  \includegraphics [width=0.5\hsize]{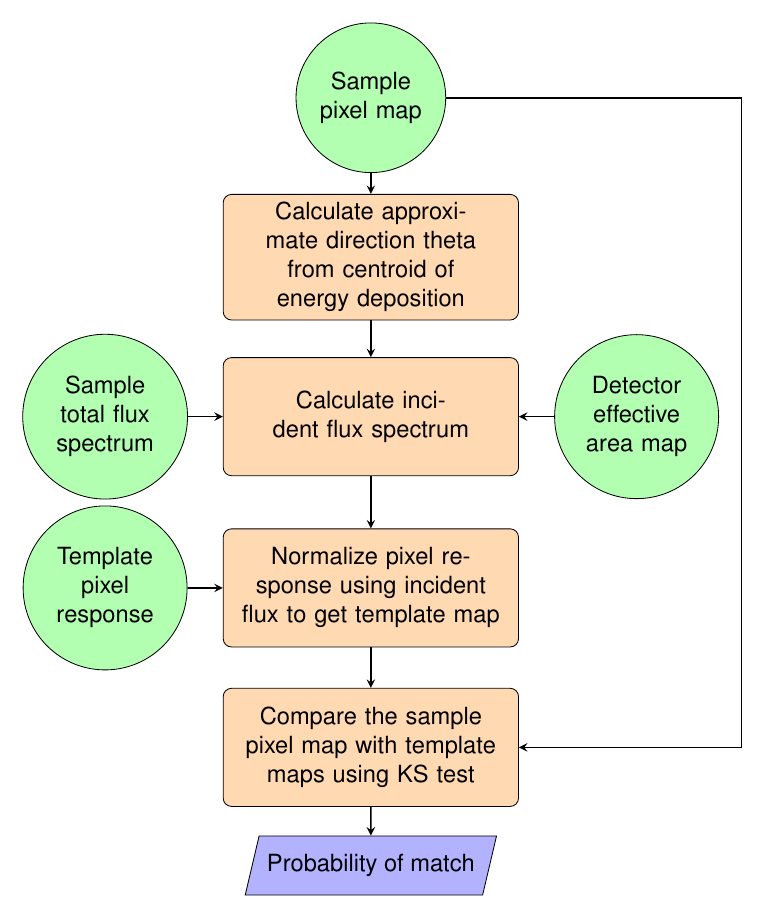}
  \caption{Flowchart of the direction reconstruction algorithm.}
  \label{fig:locflow}
\end{figure}

However, the matching probability of two pixel maps from the KS test depends on their
shapes (i.e., shape of the pixel ID vs. counts histogram) which on the other hand
depends on the source energy spectrum. So, the template pixel maps calculated beforehand
with a fixed incident energy spectrum cannot be accurately compared with the pixel map
for different sample spectrum that varies from source to source. For this purpose, we
need to re-weight the template pixel maps using the incident spectrum of the sample
source. This incident spectrum can be retrieved by unfolding the detected sample
energy spectrum by the instrument. To speed up the unfolding method algorithm, we use
an approximate method by simply dividing the detected count-rates by the effective
area of the detector in the corresponding energy bins. Thus, the approximation does not
properly reciprocate the partial energy-depositions, but works fine for the fully
contained events. The energy response function of all the pixels due to the templates
are required for the re-weighting process. The re-normalized template pixel maps are
then compared with the sample pixel maps using the KS test. The overall direction
reconstruction algorithm is outlined as a flowchart shown in Fig.~\ref{fig:locflow}.

\begin{figure}[htp]
  \centering
  \includegraphics [width=0.5\hsize]{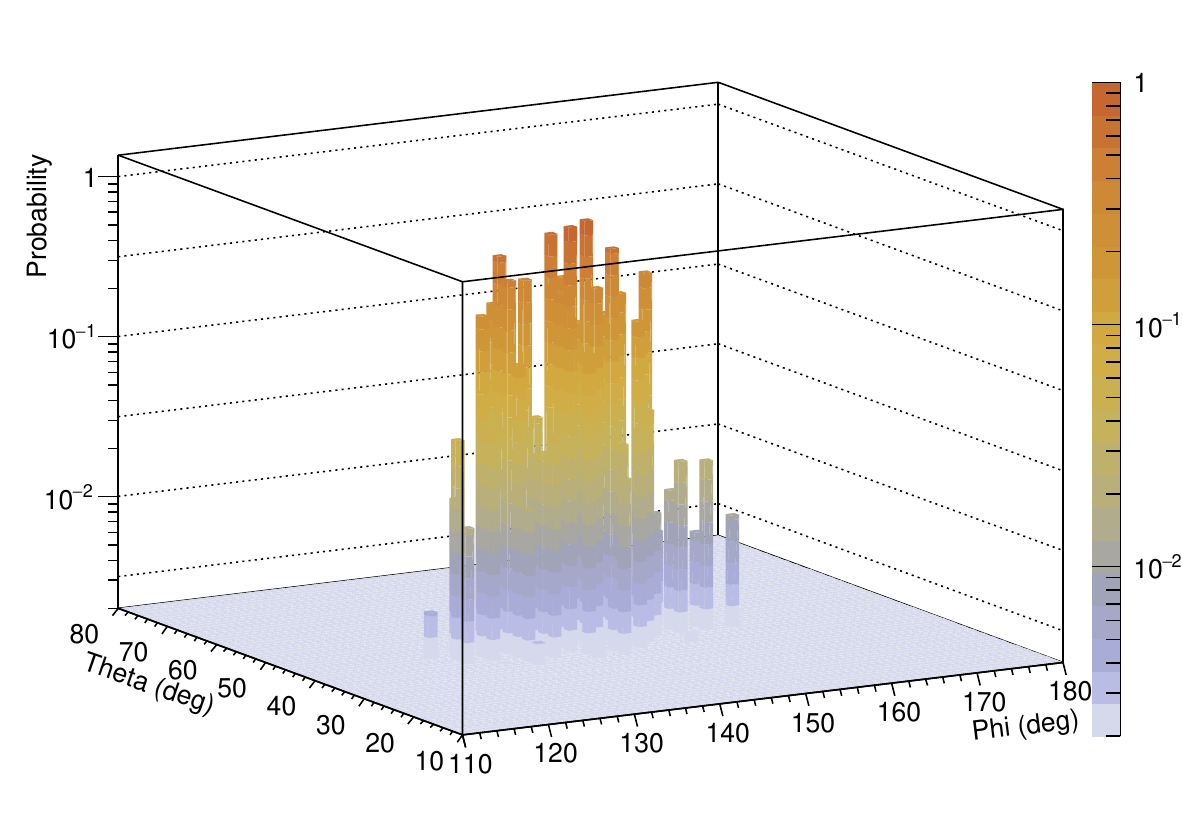}
  \caption{Matching probability distribution of a sample source resulting from the
  localization algorithm. For this example, the real location of the sample is at
  $\theta$ = 42.83$^{\circ}$, $\phi$ = 139.50$^{\circ}$, and the templates are located
  at 2$^{\circ}$ apart.}
  \label{fig:probmap}
\end{figure}

An example of the probability distribution from the direction reconstruction of an average
GRB source (like the one shown in the left panel of Fig.~\ref{fig:snr} for the Comptonized
function) assuming an arbitrary location of origin at $\theta$ = 42.83$^{\circ}$, $\phi$ =
139.50$^{\circ}$ is shown in Fig.~\ref{fig:probmap}. The discreet locations of the
templates produce this probability density forest in the vicinity of the real location.
The final reconstructed direction is obtained by taking the weighted average of the
distribution shown in Fig.~\ref{fig:probmap}. The direction reconstruction using the
templates at 2$^{\circ}$ apart gives the final $\theta$ = 42.28$^{\circ}$, $\phi$ =
140.25$^{\circ}$, and takes 5.421 s user time (0.385 s system time) for the computation.
Whereas, using the template set at 5$^{\circ}$ apart gives reconstructed $\theta$
= 42.03$^{\circ}$, $\phi$ = 142.19$^{\circ}$, and takes 0.961 s user time (0.125 s system
time) to complete the calculation. The calculation was done on a 2.8 GHz intel i7 CPU
(x86\_64 architecture) and 32 GB of RAM.

\begin{figure}[htp]
  \centering
  \includegraphics [width=0.5\hsize]{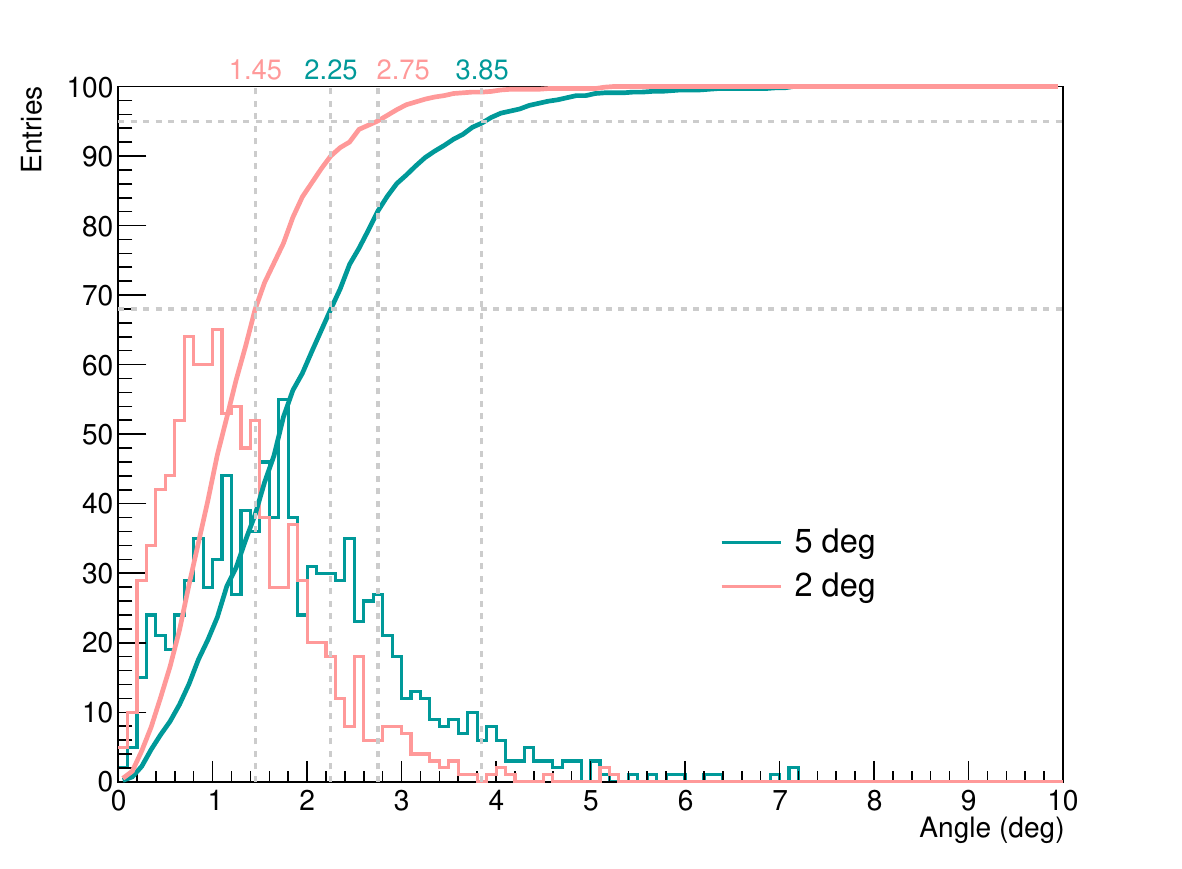}
  \caption{Angular deviation distribution for direction reconstruction shown by the
  histograms. The uncertainties of direction reconstruction at different C.L., obtained
  by integration of the histograms over different deviation angles are shown by the
  continuous lines. Results shown for both 5$^{\circ}$ and 2$^{\circ}$ template
  separation angles. The dashed gray lines and the numbers at the top mark the 68\%
  and 95\% confidences.}
  \label{fig:locuncer}
\end{figure}

\begin{figure}[htp]
  \centering
  \includegraphics [width=0.5\hsize]{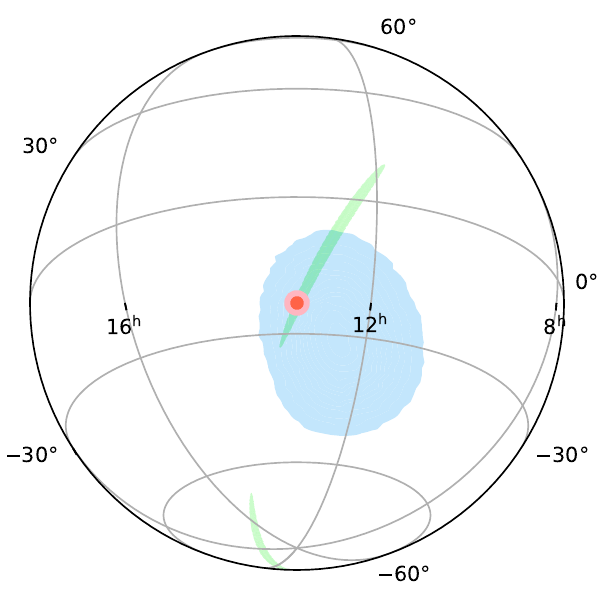}
  \caption{Skymap showing Fermi-GBM and LIGO 90\% C.L. localization regions for
  GRB170817A/GW170817 \citep{abbott_2017, abbott_2017_b, goldstein_2017}, along with 68\%
  (darker red) and 95\% (lighter red) C.L. localization uncertainty constraints
  corresponding to CE prediction.}
  \label{fig:locmap}
\end{figure}

The uncertainty of location reconstruction is calculated using $\sim$ 1000 independent
sample sources simulated from random locations uniformly distributed in the sky
hemisphere. The distributions of the deviation angles (between the actual direction
of simulation and the reconstructed direction) are shown in Fig.~\ref{fig:locuncer}, both for
5$^{\circ}$ and 2$^{\circ}$ template separation angles. To calculate the uncertainty of
direction reconstruction at different C.L. we took the running integration of the
deviation angle distribution (starting from 0$^\circ$ and normalized the integrated
value to maximum at 100); the integrated distributions are also shown in the same plot.
The calculation reveals that the location of the transients in the sky can be predicted
inside a region of radius 1.45$^{\circ}$ with 68\% confidence, and for 95\% confidence
the radius is 2.75$^{\circ}$, while using the template set of 2$^{\circ}$ separation.
The corresponding values for 5$^{\circ}$ template separation are also mentioned in
Fig.~\ref{fig:locuncer}. Furthermore, we extended the calculation using template
sources about 1$^{\circ}$ apart, which gives no further significant improvement in
the uncertainty of the prediction. The localization capability of CE is shown by the skymap
in Fig.~\ref{fig:locmap} with the example of GRB170817A. The sky locations of the same object
given by Fermi-GBM and its GW counterpart (GW170817) by rapid LIGO localization are also shown
for reference. Currently, work is ongoing to modify the localization procedure implemented
by a lightweight convolutional neural network (CNN) software, optimized for onboard execution,
to provide real-time direction reconstruction of transient events and to enable automated
low-latency alert transmission.

\section{Conclusions and outlook}
\label{sec:conc}

An all-sky instrument such as CE, capable of monitoring the sky in the hard X-ray
and low-energy $\gamma$-ray region with real-time onboard localization, will play a key
role in supporting the fast-evolving field of multi-messenger astronomy. The
combination of competitive sensitivity, wide FoV, and low-latency alert dissemination
is essential to enable rapid EM follow-up of high-energy transients and gravitational-wave
counterparts. A complete understanding of the detector response in the operational
environment is crucial to optimize the design and estimate its capabilities. In this
study, detailed simulations are performed in order to understand the detector
response to both the background and the source signals. However, the detector geometry
used in this work is a conceptual design. Optimization of the design in terms of
performance and practicality is currently ongoing, considering the feasible structural
form capable of holding the crystals and veto layers with proper dimensions and spacings
and the possibility to assemble the module with the satellite body. Performance parameters,
namely, the effective area, the sensitivity, and localization capability of the instrument,
estimated from the simulation studies, are presented here. The background environment
is a crucial aspect for estimating the performance of space-based detectors. CE orbit
having $\sim$ 20$^{\circ}$ inclination avoids hostile areas in the polar regions, where
the flux of charged particles is expected to be considerably higher. However, it will
still transit through part of the SAA region, which will affect the overall performance,
especially in terms of duty cycle.

The calculation shows that the effect of intrinsic background due to natural radioactivity
in the LYSO scintillator crystals is a crucial issue for the GRB detection in the
sub-MeV energy range, which should be addressed carefully either by selecting alternative
detector material or by some innovative technique for background suppression, like
involving machine learning techniques, which is yet to be explored. Thus, the work also
helps to choose the better configuration of the instrument taking into consideration
other practical limitations such as the budget for the experiment.

One of the most remarkable features of the CE detector is its quasi-uniform efficiency
response throughout the sky (see Fig.~\ref{fig:effarea}) due to the dome-like design
with homogeneous and granular pixels distribution. This ensures a good response of the
detector across its FoV, a feature that is particularly important for short-duration
transient monitoring. The detector is expected to perform observations for both transient
and persistent sources. Although not exclusively designed for it, it is interesting to
see the performance of the detector for $\gamma$-ray observations of persistent sources
in the keV--MeV, a range that is not extensively covered and contains multiple interesting
phenomena. However, the sensitivity levels for persistent sources are highly dependent
on the aperture or acceptance of the detector and other effects, like the long-term gain
stability of the detector. Therefore, a detailed estimation of persistent source detection
in terms of sensitivity, localization, and resolution is required and will be reported in
future work. In case of transient detection, the sensitivity threshold is consistently
above a conservative S/N = 5 for a generic GRB with fiducial spectral parameters and
integrated flux approximately above 0.6 photons\,cm$^{-2}$\,s$^{-1}$ in the 0.03--10 MeV
range. CE is comparable and, in general, has lower detection threshold fluxes than those
reported by AMEGO-X in a similar energy range for a fiducial GRB model \citep{martinez_2022}.
Although it is clear from Fig.~\ref{fig:snr} that CE (in its LYSO\,+\,GAGG configuration)
is capable of the detection of all GRBs reported by Fermi-GBM with good significance
(S/N > 3), an independent estimation of the GRB detection rate can be performed and
will be reported in the future.

It is also worth mentioning that along with the study of the transient and persistent
sources in $\gamma$ rays, the detector can also be useful to gather information about
the low-energy cosmic ray or trapped particles in the few keV to tens of MeV energy range.
This can be done by setting the provision for different trigger logics in the onboard
trigger configuration of the detector to acquire the background data. This additional
use of the detector can provide important information in the low-energy cosmic ray study.

A source localization algorithm is developed to promptly estimate the precise position
of possible transient outbursts across the detector FoV. The analysis
performed using an average GRB spectral model leads to a 95\% confidence region of
$\sim$ 2.75$^\circ$ radius ($\sim$ 1.45$^\circ$ for 68\% C.L.), suggesting that the
instrument has an online localization precision that is better by about an order of
magnitude than those typically reported by monitors like Fermi-GBM for this type of
event. However, in this study, we considered an average type of GRB for the
calculation, which in principle can be done by using a variety of GRB spectra, to obtain a
more general idea for localization precision. A more advanced localization algorithm,
based on a lightweight CNN, is currently being implemented for fully autonomous onboard
operation, enabling real-time transient localization without the need for ground-based
processing. Crystal~Eye will therefore complement the global effort toward rapid and
coordinated multi-messenger astronomy, providing real-time high-energy transient
localization and enabling timely EM follow-up across the widest possible range of
observatories.

\section*{Authors contribution}
\label{credit}

All authors are listed alphabetically according to the Crystal~Eye collaboration agreement.
{\bf R.~Sarkar:} Formal analysis, Investigation, Conceptualization, Methodology, Software,
Visualization, Writing – original draft, Writing – review and editing
{\bf F.C.T.~Barbato:} Funding acquisition, Conceptualization, Supervision, Project administration,
Writing – review and editing
{\bf G.~Oganesyan:} Resources, Writing – review and editing
{\bf L.~Wu:} Software, Writing – review and editing
{\bf M.~Fernandez Alonso:} Software, Writing – review and editing
{\bf Others:} Writing – review and editing

\section*{Acknowledgements}
\label{ackn}

This work has been funded by the European Union -- NextGenerationEU, Mission 4,
Component 2, under the Italian Ministry of University and Research (MUR) National
Innovation Ecosystem grant ECS00000041 -- VITALITY -- CUP D13C21000430001.
Also funded by the European Union -- NextGenerationEU, National Recovery And
Resilience Plan (NRRP) -- Mission 4 Component 2 Investment 1.1 -- ``Fund For The
National Research Program And For Projects of National Interest (NRP)'' -- 
``WINK: a pathfinder mission for the future Crystal Eye X and gamma rays all
sky monitor'', bando D.D. MUR n. 104/2022 --  CUP: D53D23002550006.
AT and FC acknowledge financial support from the Istituto Nazionale di Astrofisica
(INAF) through the grant 1.05.23.05.06. 
FS acknowledges financial support from the AHEAD2020 project (grant agreement 
n. 871158). BB and MB acknowledge financial support from the Italian Ministry of
University and Research (MUR) for the PRIN grant METE under contract no. 2020KB33TP.
  

\bibliographystyle{elsarticle-harv} 
\bibliography{references}

\end{document}